\documentclass[10pt,twoside]{article}
\usepackage{amssymb}

\usepackage{graphicx}

\begin{document}

\title{Bayesian Statistics and Parameter Constraints on the\\
Generalized Chaplygin Gas Model using SNe Ia Data}
\author{R. Colistete Jr.\thanks{%
e-mail: \texttt{colistete@cce.ufes.br}}, J. C. Fabris\thanks{%
e-mail: \texttt{fabris@cce.ufes.br}} and S.V.B. Gon\c{c}alves\thanks{%
e-mail: \texttt{sergio@cce.ufes.br}} \\
\\
\mbox{\small Universidade Federal do Esp\'{\i}rito Santo,
Departamento
de F\'{\i}sica}\\
\mbox{\small Av. Fernando Ferrari s/n - Campus de Goiabeiras, CEP
29060-900, Vit\'oria, Esp\'{\i}rito Santo, Brazil}}
\date{\today}
\maketitle

\begin{abstract}
The type Ia supernovae (SNe Ia) observational data are used to estimate the
parameters of a cosmological model with cold dark matter and the generalized
Chaplygin gas model (GCGM). The GCGM depends essentially on five parameters:
the Hubble constant, the parameter $\bar{A}$ related to the velocity of the
sound, the equation of state parameter $\alpha$, the curvature of the
Universe and the fraction density of the generalized Chaplygin gas (or the
cold dark matter). The parameter $\alpha$ is allowed to take negative values
and to be greater than $1$. The Bayesian parameter estimation yields $\alpha
= - 0.86^{+6.01}_{-0.15}$, $H_0 = 62.0^{+1.32}_{-1.42}\,km/Mpc.s$, $\Omega
_{k0}=-1.26_{-1.42}^{+1.32}$, $\Omega_{m0} = 0.00^{+0.86}_{-0.00}$, $%
\Omega_{c0} = 1.39^{+1.21}_{-1.25}$, $\bar A =1.00^{+0.00}_{-0.39}$, $t_0 =
15.3^{+4.2}_{-3.2}$ and $q_0 = -0.80^{+0.86}_{-0.62}$, where $t_0$ is the
age of the Universe and $q_0$ is the value of the deceleration parameter
today. Our results indicate that a Universe completely dominated by the
generalized Chaplygin gas is favoured, what reinforces the idea that the
this gas may unify the description for dark matter and dark energy, at least
as the SNe Ia data is concerned. A closed and accelerating Universe is also
favoured. The traditional Chaplygin gas model (CGM), $\alpha = 1$ is not
ruled out, even if it does not give the best-fitting. Particular cases with
four or three independent free parameters are also analysed. \vspace{0.7cm}
\end{abstract}

PACS number(s): 98.80.Bp, 98.80.Es, 04.60.Gw \vspace{0.7cm}

\section{Introduction}

One of the most important problems today in cosmology is the nature of dark
matter and dark energy that must dominate the matter content of the Universe
\cite{sahni}. The existence of dark matter is suggested by the anomalies in
the dynamics of galaxies and clusters of galaxies \cite{galaxies}. Dark
energy seems to be an inevitable consequence of the present acceleration of
the Universe as indicated by the type Ia supernovae (SNe Ia) data, which
asks for a fluid of negative pressure which does not agglomerate at small
scales \cite{riess1998,perlmutter}. While in general the nature of dark
matter can be connected with relics of fundamental theories, like axions
\cite{axions}, there are two main candidates to represent dark energy:
cosmological constant \cite{weinberg} and quintessence which is a self
interacting scalar field \cite{steinhardt}--\cite{brax}. Both models suffers
of many drawbacks, mainly linked with fine tuning of parameters, as in the
quintessence model \cite{lyth}, or disagreement with the theoretically
predicted values, like in the cosmological constant case \cite{carroll}.

The traditional Chaplygin gas model (CGM) and the\ generalized Chaplygin gas
model (GCGM) have been widely considered as alternatives to the cosmological
constant and to quintessence as the dark energy that drives the present
acceleration of the Universe \cite{pasquier}--\cite{bento2002}. The
appealing of the GCGM comes from, among other reasons, the fact that it can
unify the description of dark matter and dark energy. The generalized
Chaplygin gas (GCG) is characterized by the equation of state
\begin{equation}
p=-\frac{A}{\rho ^{\alpha }}\quad ,  \label{eos}
\end{equation}
where $A$ and $\alpha $ are constants that in general are considered as
positive numbers. The conservation equation leads to the relation
\begin{equation}
\rho =\biggr\{A+\frac{B}{a^{3(1+\alpha )}}\biggl\}^{1/(1+\alpha )}\quad .
\label{densidade}
\end{equation}
Some of the main interesting features of eq. (\ref{densidade}) are the
following. Initially, the GCGM behaves as a dust fluid, with $\rho \propto
a^{-3}$, while at late times the GCGM behaves as a cosmological constant
term, $\rho \propto A^{1/(1+\alpha )}$. Hence, the GCGM interpolates a
matter dominated phase (where the formation of structure can happen) and a
de Sitter phase. At the same time, the sound velocity of the GCGM is
positive which assures that, in spite of exhibiting a negative pressure, it
is stable against perturbations of the background configuration \cite{jerome}%
.

A lot of effort has been done in order to constrain the free parameters of
the GCGM. Type Ia supernovae data \cite{roberto}--\cite{biesiada}, the
spectra of anisotropy of cosmic microwave background radiation \cite
{bertolami2004a}, the mass power spectrum \cite{bento2004,bilic2003,fabris},
gravitational lenses \cite{silva}, X-ray data \cite{cunha} and age estimates
of high-$z$ objects \cite{alcaniz2003b} have extensively been used in this
sense. This is essential for many obvious reasons. One of them is that if
the parameter $\alpha $ is close to zero, the GCGM reduces essentially to
the cosmological constant model. If observations indicate that this is the
preferred value, the appealing for the cosmological constant as dark energy
is reinforced. In a previous work \cite{roberto}, we have analysed the case
of the traditional Chaplygin gas model, where $\alpha =1$, using the SNe Ia
data. One important aspect of this analysis is the use of the Bayesian
statistics, for one side, and also the fact that all free parameters (Hubble
constant, the curvature term, the value of the sound velocity and the
proportion of ordinary matter and Chaplygin gas) of the model were taken
into account in the treatment of the problem. This contrasts with most of
the analyses previously found in the literature in the sense that they
employ a more simplified statistical analysis or some input parameters were
fixed a priori \cite{martin}--\cite{biesiada}. In that work \cite{roberto},
it was concluded that the unification scenario (i.e., without pressureless
matter), and a sound velocity near (but not equal) to the velocity of light
were preferred. Moreover, a closed Universe was clearly favoured.

The aim of the present work is to apply the same analysis, using
the Bayesian statistics and taking into account all free
parameters, to the GCGM. In order to keep contact with the
previous work, we will use the same supernovae sample. This sample
is much smaller than those available today (see, for example, ref.
\cite{riess2004}--\cite{barris}). But, the quality of the data is
extremely good, which leads to a smaller value of $\chi _{\nu
}^{2}$ (the quantity that measures the quality of the fitting):
with this restricted sample the best-fitting gives $\chi _{\nu
}^{2}$ around $0.74$, while for the complete ``gold'' sample
\cite{riess2004}, the value mounts to unity. In the GCGM there are
five free parameters instead of the four free parameters of the
traditional Chaplygin gas model ($\alpha =1$): the Hubble constant
$H_{0}$, the parameter $\bar{A}$ related to the sound velocity of
the Chaplygin gas, the curvature density parameter $\Omega _{k0}$,
the ordinary matter parameter $\Omega _{m0}$ (or alternatively,
the Chaplygin gas parameter $\Omega _{c0}$) and the equation of
state parameter $\alpha $.

One important point in performing this generalized analysis is the range of
validity of the parameter $\alpha $. Usually, it has been assumed that $%
0\leq \alpha \leq 1$. The reason is twofold: $\alpha $ greater than one
could lead, even if in the future, to a sound velocity greater than the
velocity of the light; $\alpha $ negative should lead to an imaginary sound
velocity, that is, instability. However, an analysis of the supernovae data
may taken into account values of $\alpha $ greater than one or negative. In
particular, in reference \cite{bertolami2004b}, using a different
statistical approach with respect to the method employed here, a value of $%
\alpha $ near $2$ was favoured. In fact, values of $\alpha $ outside that
limited range can be considered. The reason lies in the fact that the GCGM
with the equation of state (\ref{eos}) may be a phenomenological
manifestation of a theory expressed in terms, for example, of scalar field
with a potential, or even of tachyon condensate model \cite{gorini, chimento}%
. In this case, the notion of sound velocity can not be identified with the
ordinary expression used in the perfect fluid approach, and no violation of
causality or instability may occur if $\alpha $ is greater than one or
negative.

Hence, in our analysis the parameter $\alpha $ will vary from negative to
highly positive values. The prediction for each parameter ($\alpha $, $H_{0}$%
, $\Omega _{k0}$, $\Omega _{m0}$, $\Omega _{c0}$ and $\bar{A}$) is taken
marginalizing, i.e., integrating over the other parameters. We stress that
such marginalization procedure gives predictions that can be quite different
from a two dimensional analysis, for example, where one searches for the
simultaneous preferred values for any two parameters. This marginalization
is in fact essential in order to have the correct prediction for a given
free parameter. We will consider the following cases: all five parameters
free; spatial curvature zero (four free parameters); pressureless matter
given only by the barionic component (four free parameters); pressureless
matter absent (four free parameters); spatial curvature zero and no
pressureless matter component (three free parameters); spatial curvature
zero and pressureless matter component given by the barionic component
(three free parameteres). For the more general case with five free
parameters the predictions we obtain are the following: $\alpha
=-0.86_{-0.15}^{+6.01}$, $H_{0}=62.0_{-1.42}^{+1.32}\,km/Mpc.s$, $\Omega
_{k0}=-1.26_{-1.42}^{+1.32}$, $\Omega _{m0}=0.00_{-0.00}^{+0.86}$, $\Omega
_{c0}=1.39_{-1.25}^{+1.21}$, $\bar{A}=1.00_{-0.39}^{+0.00}$. Moreover the
predicted age of the Universe and the value of de deceleration parameter
today are $t_{0}=15.3_{-3.2}^{+4.2}$ and $q_{0}=-0.80_{-0.62}^{+0.86}$,
respectively. The results are consistent with, for example, the age of
globular clusters \cite{krauss}. Two crucial parameters are $\alpha $ and $%
\bar{A}$, since if $\alpha =0$ and/or $\bar{A}=1$ are favoured,
the GCGM becomes equivalent to the cosmological constant model
($\Lambda $CDM). The value of the parameter $\bar{A}$ is in fact
peaked at unity. But for reasons discussed later, this does not
imply that this is the most favoured value; indeed it is just
expected to be near unity. On the other hand, the parameter
$\alpha $ is peaked in a negative value, but the spread is very
large. In particular, the traditional Chaplygin gas with $\alpha
=1$ can not be excluded or even disfavoured: in most of the cases
this value of $\alpha $ is favoured with a probability around
$50\%$. Hence, our analysis does not exclude the traditional
Chaplygin gas model.

This paper is organized as follows. In next section, the basic
equations and quantities are exhibited. Section $3$ explains the
type Ia supernovae fitting. In section $4$, we display the results
of the Bayesian analysis of type Ia supernovae data with the GCGM.
We present our conclusions in section $5$. \newline

\section{Description of the GCG model}

Regardless the possibility of unification of dark matter and dark energy
using the GCGM, we will consider a model for the Universe today with two
fluids: pressureless matter and the GCG. In doing so, our aim is to verify
whether the unified model is the case favoured by observations. Moreover,
even if in the GCGM there is no need of a dark matter component, there is
baryonic matter in the Universe which is represented also by a pressureless
matter. Hence, the dynamics of the Universe is driven by the Friedmann's
equation
\begin{eqnarray}
\biggr(\frac{\dot{a}}{a}\biggl)^{2}+\frac{k}{a^{2}} &=&\frac{8\pi G}{3}%
\biggr(\rho _{m}+\rho _{c}\biggl)\quad ,  \label{be3} \\
\dot{\rho}_{m}+3\frac{\dot{a}}{a}\rho _{m} &=&0\quad ,  \label{be4} \\
\dot{\rho}_{c}+3\frac{\dot{a}}{a}\biggr(\rho _{c}-\frac{A}{\rho _{c}^{\alpha
}}\biggl) &=&0\quad ,  \label{be5}
\end{eqnarray}
where $\rho _{m}$ and $\rho _{c}$ stand for the pressureless matter and
generalized Chaplygin gas component. Dot means derivative with respect to
the cosmic time $t$. As usual, $k=0,1,-1$ indicates a flat, closed and open
spatial section. In the conservation equation for the generalized Chaplygin
gas component (\ref{be5}), the equation of state (\ref{eos}) has been
introduced.

The equations expressing the conservation law for each fluid (\ref{be4})-(%
\ref{be5}) lead to
\begin{equation}
\rho _{m}=\frac{\rho _{m0}}{a^{3}}\quad ,\quad \rho _{c}=\biggr\{A+\frac{B}{%
a^{3(1+\alpha )}}\biggl\}^{1/(1+\alpha )}\quad ,
\end{equation}
Henceforth, we will parametrize the scale factor such that its value today
is equal to unity, $a_{0}=1$. Hence, $\rho _{m0}$ and $\rho _{c0}=\biggr\{A+B%
\biggl\}^{1/(1+\alpha )}$ are the pressureless matter and GCG densities
today. Eliminating from the last relation the parameter $B$, the GCG density
at any time can be reexpressed as
\begin{equation}
\rho _{c}=\rho _{c0}\biggr\{\bar{A}+\frac{1-\bar{A}}{a^{3(1+\alpha )}}%
\biggl\}^{1/(1+\alpha )}\quad ,
\end{equation}
where $\bar{A}=A/\rho _{c0}$. This parameter $\bar{A}$ is connected with the
sound velocity for the GCG today by the relation $v_{s}^{2}=\alpha \bar{A}$.

The luminosity distance is given by the expression
\begin{equation}
D_{L}=\frac{a_{0}^{2}}{a}r_{1}\quad ,
\end{equation}
$r_{1}$ being the co-moving coordinate of the source. Using the expression
for the propagation of light
\begin{equation}
ds^{2}=0=dt^{2}-\frac{a^{2}dr^{2}}{1-kr^{2}}\quad ,
\end{equation}
and the Friedmann's equation (\ref{be3}), we can reexpress the luminosity
distance as
\begin{equation}
D_{L}=(1+z)S[f(z)]\quad ,
\end{equation}
where
\begin{equation}
S(x)=x\quad (k=0),\quad S(x)=\sin x\quad (k=1),\quad S(x)=\sinh x\quad
(k=-1)\quad ,
\end{equation}
and
\begin{equation}
f(z)=\frac{1}{H_{0}}\int_{0}^{z}\frac{d\,z^{\prime }}{\{\Omega
_{m0}(z^{\prime }+1)^{3}+\Omega _{c0}[\bar{A}+(z^{\prime }+1)^{3(1+\alpha
)}(1-\bar{A})]^{1/(1+\alpha )}-\Omega _{k0}(z^{\prime }+1)^{2}\}^{1/2}}\quad
,
\end{equation}
with the definitions
\begin{equation}
\Omega _{m0}=\frac{8\pi G}{3}\frac{\rho _{m0}}{H_{0}^{2}}\quad ,\quad \Omega
_{c0}=\frac{8\pi G}{3}\frac{\rho _{c0}}{H_{0}^{2}}\quad ,\quad \Omega _{k0}=-%
\frac{k}{H_{0}^{2}}\quad ,
\end{equation}
where $\Omega _{m0}+\Omega _{c0}+\Omega _{k0}=1$. The final equations have
been also expressed in terms of the redshift $z=-1+\frac{1}{a}$.

The age of the Universe today can be expressed as
\begin{equation}
\frac{t_{0}}{T}=\int_{0}^{z}\frac{d\,z^{\prime }}{(1+z^{\prime })\{\Omega
_{m0}(z^{\prime }+1)^{3}+\Omega _{c0}[\bar{A}+(z^{\prime }+1)^{3(1+\alpha
)}(1-\bar{A})]^{1/(1+\alpha )}-\Omega _{k0}(z^{\prime }+1)^{2}\}^{1/2}}\quad
,
\end{equation}
where $T=(100/H_{0})\times 10^{10}$, so $t_{0}$ has units of years.

The value of the decelerated parameter today, $q_{0}=-\frac{a\ddot{a}}{\dot{a%
}^{2}}$, is given by
\begin{equation}
q_{0}=\frac{\Omega _{m0}+\Omega _{c0}(1-3\bar{A})}{2}\quad .
\end{equation}

The value $a_{i}$\ of the scale factor that shows the beginning of the
recent accelerating phase of the Universe is given by the roots of the
equation
\begin{equation}
\ddot{a}=a(\dot{H}+H^{2})\quad ,
\end{equation}
and is related to the redshift value $z_{i}$ such that $\frac{a_{i}}{a_{0}}=%
\frac{1}{1+z_{i}}$ or $z_{i}=-1+\frac{a_{0}}{a_{i}}$.

\section{Supernovae fitting}

We proceed by fitting the SNe Ia data using the GCG model described above.
Essentially, we compute the quantity distance moduli,
\begin{equation}
\mu _{0}=5\log \left( \frac{D_{L}}{M\!pc}\right) +25\quad ,
\end{equation}
and compare the same distance moduli as obtained from observations. The
quality of the fitting is characterized by the $\chi ^{2}$ parameter of the
least-squares statistic, as defined in Ref. \cite{riess1998},
\begin{equation}
\chi ^{2}=\sum_{i}\frac{\left( \mu _{0,i}^{o}-\mu _{0,i}^{t}\right) ^{2}}{%
\sigma _{\mu _{0},i}^{2}+\sigma _{mz,i}^{2}}\quad .  \label{Chi2}
\end{equation}
In this expression, $\mu _{0,i}^{o}$ is the measured value, $\mu _{0,i}^{t}$
is the value calculated through the model described above, $\sigma _{\mu
_{0},i}^{2}$ is the measurement error, $\sigma _{mz,i}^{2}$ is the
dispersion in the distance modulus due to the dispersion in galaxy redshift
due to peculiar velocities. This quantity we will taken as
\begin{equation}
\sigma _{mz}=\frac{\partial \log D_{L}}{\partial z}\sigma _{z}\quad ,
\end{equation}
where, following Ref. \cite{riess1998,wang}, $\sigma _{z}=200\,km/s$.

In this article, we have used the same $26$ type Ia supernovae in table $1$
of ref. \cite{roberto}, so we do not need to analyse and \ compare again the
CGM with the $\Lambda $CDM (this work has been done in ref. \cite{roberto}),
but instead we can concentrate our efforts on the GCGM. Although nowadays
there are approximately $200$ SNe Ia available, the choice of these $26$ are
confirmed to be of excellent quality by analysing the low values of $\chi
_{\nu }^{2}$ (the estimated errors for degree of freedom, i.e., $\chi ^{2}$
divided by $26$, the number of SNe Ia), shown in table \ref{tableBestFitGCG}%
. $\chi _{\nu }^{2}$ values range between $0.738$ to $0.748$, while other
larger SNe Ia samples have been used in many cosmological models \cite
{riess2004, tonry, nesseris}, typically resulting $\chi _{\nu }^{2}$ near
the unity.

\begin{table}[!t]
\begin{center}
\begin{tabular}{|c|c|c|c|c|c|c|}
\hline\hline
&  &  &  &  &  &  \\[-7pt]
& GCGM & GCGM : & GCGM : & GCGM : & GCGM : $k=0$, & GCGM : $k=0$, \\
&  & $k=0$ & $\Omega_{m0}=0$ & $\Omega_{m0}=0.04$ & $\Omega_{m0}=0$ & $%
\Omega_{m0}=0.04$ \\[2pt] \hline
&  &  &  &  &  &  \\[-7pt]
$\chi_{\nu }^{2}$ & $0.7383$ & $0.7386$ & $0.7429$ & $0.7412$ & $0.7475$ & $%
0.7441$ \\[2pt] \hline
&  &  &  &  &  &  \\[-7pt]
$\alpha$ & $42.04$ & $42.98$ & $50.36$ & $46.63$ & $13.00$ & $27.00$ \\%
[2pt] \hline
&  &  &  &  &  &  \\[-7pt]
$H_{0}$ & $62.4$ & $62.3$ & $62.4$ & $62.4$ & $62.8$ & $62.7$ \\[2pt] \hline
&  &  &  &  &  &  \\[-7pt]
$\Omega_{k0}$ & $-0.08$ & $0$ & $0.18$ & $0.13$ & $0$ & $0$ \\[2pt] \hline
&  &  &  &  &  &  \\[-7pt]
$\Omega_{m0}$ & $0.167$ & $0.131$ & $0$ & $0.04$ & $0$ & $0.04$ \\%
[2pt] \hline
&  &  &  &  &  &  \\[-7pt]
$\Omega_{c0}$ & $0.91$ & $0.87$ & $0.82$ & $0.83$ & $1$ & $0.96$ \\%
[2pt] \hline
&  &  &  &  &  &  \\[-7pt]
$1-\bar{A}$ & $2.2 \times 10^{-16}$ & $2.2 \times 10^{-16}$ & $3.3 \times
10^{-16}$ & $5.5 \times 10^{-16}$ & $1.3 \times 10^{-4}$ & $5.5 \times
10^{-9}$ \\[2pt] \hline
&  &  &  &  &  &  \\[-7pt]
$t_{0}$ & $14.0$ & $14.1$ & $14.4$ & $14.3$ & $14.0$ & $14.0$ \\[2pt] \hline
&  &  &  &  &  &  \\[-7pt]
$q_{0}$ & $-0.83$ & $-0.80$ & $-0.82$ & $-0.82$ & $-1.00$ & $-0.94$ \\%
[2pt] \hline
&  &  &  &  &  &  \\[-7pt]
$a_{i}$ & $0.756$ & $0.760$ & $0.790$ & $0.779$ & $0.795$ & $0.792$ \\%
[2pt] \hline
&  &  &  &  &  &  \\[-7pt]
$m$ & $-0.36$ & $-0.36$ & $-0.30$ & $-0.32$ & $-0.27$ & $-0.29$ \\%
[2pt] \hline
&  &  &  &  &  &  \\[-7pt]
$n$ & $-0.34$ & $-0.21$ & $-0.14$ & $-0.23$ & $-0.29$ & $-0.26$ \\%
[2pt] \hline\hline
\end{tabular}
\end{center}
\caption{The best-fitting parameters, i.e., when $\protect\chi _{\protect\nu
}^{2}$ is minimum, for each type of spatial section and matter content of
the generalized Chaplygin gas model. $H_{0}$ is given in $km/M\!pc.s$, $\bar{%
A}$ in units of $c$, $t_{0}$ in $Gy$ and $a_{i}$ in units of $a_{0}$. $m$
and $n$ are the constants of the relation $log(1-\bar{A})=m\,\protect\alpha %
+n$. }
\label{tableBestFitGCG}
\end{table}

The best-fitting independent parameters and other dependent
parameters (functions of the independent parameters) for the
generalized Chaplygin gas model (GCGM) are given by table
\ref{tableBestFitGCG}, for six different cases of spatial section
and matter content. The first case (GCGM) has five
free independent parameters $(\alpha ,H_{0},\Omega _{m0},\Omega _{c0},\bar{A}%
)$, the second case ($GCGM:k=0$) has four free independent parameters $%
(\alpha ,H_{0},\Omega _{m0},\bar{A})$, the third
($GCGM:\Omega_{m0}=0$) and fourth cases ($GCGM:\Omega _{m0}=0.04$)
have four free independent parameters
$(\alpha ,H_{0},\Omega _{c0},\bar{A})$, the fifth ($GCGM:k=0,\ \Omega _{m0}=0$%
) and sixth ($GCGM:k=0,\ \Omega _{m0}=0.04$) cases have three free
independent parameters $(\alpha ,H_{0},\bar{A})$.

It is important to emphasize that, for each case, all free independent
parameters are considered simultaneously to obtain the minimum of $\chi
_{\nu }^{2}$. So, for example assuming the GCGM with $k=0$ and $\Omega
_{m0}=0$, if we ask for the best simultaneous values of $(\alpha ,H_{0},\bar{%
A})$ then the answer is given by last column of table \ref{tableBestFitGCG}.
However, in this example, asking for the best value of $\alpha $ by weighing
(marginalizing or integrating) all possible values of $(H_{0},\bar{A})$ is
another issue which is answered by the Bayesian estimations of the next
section, not by best-fitting in $n$-dimensional parameter space whose
results are listed in table \ref{tableBestFitGCG}.

It is remarkable that all independent and dependent parameters are
physically acceptable : the cold dark matter density parameter range is $%
0.13<\Omega _{m0}<0.17$ (when it is not fixed \textit{a priori}), the age of
Universe today sits between $14.0\,Gy$ and $14.4\,Gy$, the value $a_{i}$ of
the scale factor that marks the beginning of the recent accelerating phase
of the Universe goes from $0.76\,a_{0}$ to $0.80\,a_{0}$, etc.

But the values of $\alpha $ and $\bar{A}$ are somewhat a surprise : $\alpha $
is extremely large and $\bar{A}$ is as close to $1$ as $\alpha $ is large.
The minimization of $\chi _{\nu }^{2}$ was obtained in the following way :
starting with an initial global minimum taken from the $n$-dimensional
discrete parameter space (see next section), usually around $(\alpha ,\bar{A}%
)\approx (3.0,0.95)$, we increment $\alpha $ slowly at each step, the new
step uses the last step parameter values as initial values to search the
local minimum of $\chi _{\nu }^{2}$ (by using the function \textit{%
FindMinimum} of the software \textit{Mathematica} \cite{mathematica}) and so
forth. So, for each value of $\alpha $ there is a value of $\bar{A}$, the
relation between $\bar{A}$ and $\alpha $ is simple indeed : $\bar{A}%
=1-10^{m\,\alpha +n}$ or $log(1-\bar{A})=m\,\alpha +n$, with $m\approx -0.3$
and $n\approx -0.25$, or see the table \ref{tableBestFitGCG} for each case
of spatial section and matter content.

The minima of $\chi _{\nu }^{2}$ as function of $\alpha $ change very
slowly. For example, the first case (GCG) has the minimum of $\chi _{\nu
}^{2}$ starting with $0.750$ (when $\alpha =3.5$), decreasing to $0.744$
(when $\alpha =7$) and reaching the best minimum $0.738$ (when $\alpha
=42.04 $). Through all the evolution of $\alpha $, the independent and
dependent parameters have physically acceptable values, so low values of $%
\alpha $ are only barely worse than large values of $\alpha $ with respect
to $\chi _{\nu }^{2}$ minimization.

Slightly smaller values of $\chi _{\nu }^{2}$ favour the GCGM over
the CGM (compare with table $3$ of \cite{roberto}), independent of
the spatial section type and matter content. The next section
employs the Bayesian statistics to obtain a better comparison
between these models and a more robust estimation of parameters.

The \textit{Mathematica} package used for the Bayesian analyses,
developed by one of the authors (R. C. Jr.), is due to be publicly
released. It comprises functions for calculation, marginalization,
credible/confidence regions and intervals, maximization and
visualization.

\section{Bayesian analyses of the cosmological parameters}

The same Bayesian statistics approach presented in section $3$ of ref. \cite
{roberto} was employed here to obtain the parameter estimations and answers
for some hypothesis about the generalized Chaplygin gas model (GCGM) with up
to five free parameters $(\alpha ,H_{0},\Omega _{m0},\Omega _{c0},\bar{A}) $%
, the age of the Universe $t_{0}$, the deceleration parameter $q_{0}$ and
the value $a_{i}$ of the scale factor that shows the beginning of the recent
accelerating phase of the Universe. The estimations have a central value
where the one-dimensional PDF (Probability Distribution Function)\ is
maximum and the positive and negative uncertainties are defined at a $%
2\sigma $ ($95.45\%$) credible (or confidence) level, see more details in
section $3$ of ref. \cite{roberto}.

Each independent parameter estimation used an one-dimensional PDF obtained
by marginalization, i.e., where ($n-1)$-dimensional integrals are computed
for each value of the parameter in the case of a $n$-dimensional parameter
space. For example, in the $5$-parameters case (without any restriction on $%
\Omega _{k0}$ and $\Omega _{m0}$), the $p(\alpha \mid \mu _{0})$ is obtained
through marginalization process described by $4$-dimensional integrals of $%
p(\alpha ,H_{0},\Omega _{m0},\Omega _{c0},\bar{A}\mid \mu _{0})$ over a $4$%
-dimensional parameter space.

\begin{figure}[t]
\begin{center}
\includegraphics[scale=0.8]{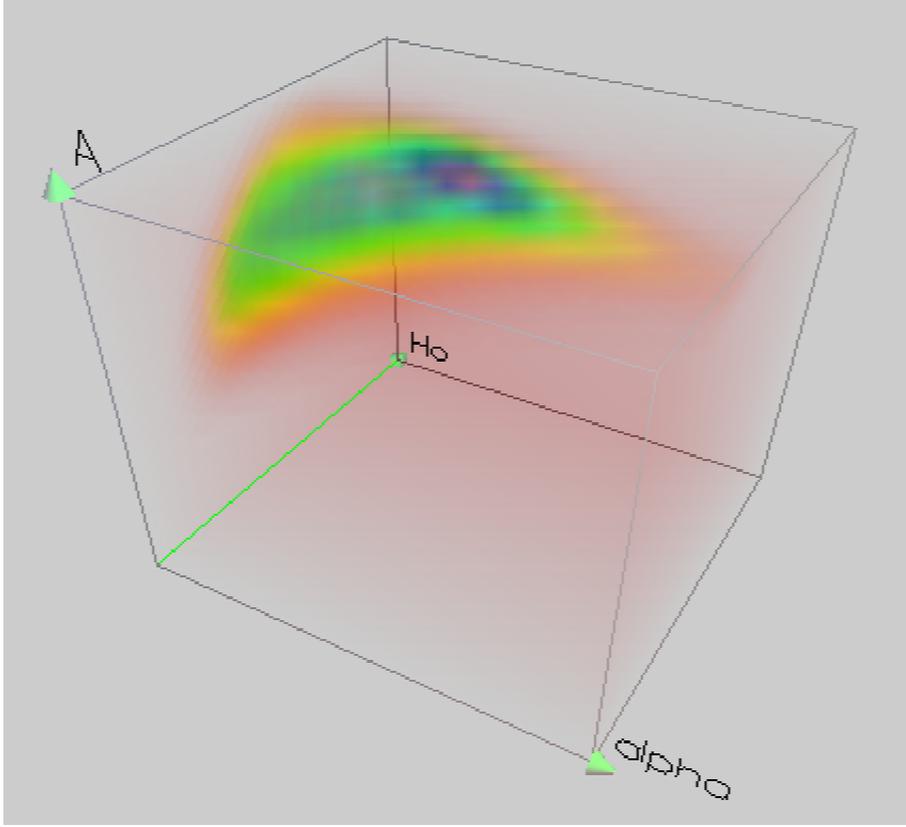}
\end{center}
\caption{{\protect\footnotesize The graphics of the joint PDF as function of
$(\protect\alpha ,H_{0},\bar{A})$ for the generalized Chaplygin gas model
when fixing\ $k=0$ and $\Omega _{m0}=0.04$. It is a $3D$ density plot where
the function (here the PDF) is rendered as semi-transparent colourful gas,
or we can say that the plot is made by voxels (volume elements) analogous to
pixels (picture elements). Where the PDF is minimum, the transparency is
total and the colour is red, where the PDF is maximum then the voxel is
opaque and the colour is violet; mid-range values are semi-transparent and
coloured between red and violet (red, orange, yellow, green, blue, violet).
The parameter ranges are $-1<\protect\alpha \leqslant 7$, $55\leqslant
H_{0}\leqslant 70$ and $0\leqslant \bar{A}\leqslant 1$, with resolutions $%
0.2\times 1\times 0.05$, respectively. The joint PDF of $(\protect\alpha
,H_{0},\bar{A})$ for other cases ($5$ free parameters, $4$ free parameters,
etc.)\ show a similar $3D$ shape. }}
\label{figPdf3Dm04}
\end{figure}

An ideal calculation to compute the $n$-dimensional integrals
would include an infinite number of samples of parameter space
points with infinite volume, but in practical estimations we are
limited to choose a finite region of the parameter space (such
that outside it the probabilities are almost null) and a finite
number of samples. With this criteria, we have chosen $-1<\alpha
\leqslant 7$ with resolution from $0.2$ to $0.5$. The lower
resolution of $0.5$ is employed in the regions where the PDF
changes slowly with respect to $\alpha $, which is interpolated to
generate a higher resolution of $0.2$ for two and one-dimensional
PDFs. Analogously, $55\leqslant H_{0}\leqslant 70$ with resolution
of\ $1$, $0\leqslant \Omega _{m0}\leqslant 1$ with resolution of\
$0.05$, $0\leqslant \Omega
_{c0}\leqslant 3.4$ with resolution of\ $0.17$ and $0\leqslant \bar{A}%
\leqslant 1$ with resolution of\ $0.05$. So there is a $5$-dimensional
discrete space of $23\times 16\times 21\times 21\times 21$ points (total of $%
3,408,048$ points whose PDF values were calculated) as well as lower
dimensional discrete spaces (fixing $\Omega _{m0}$, etc.), and the
continuous PDFs in three, two and one-dimensional spaces are obtained by
marginalizing interpolated functions built from the discrete points. We
obviously discard the parameter space regions where the PDFs are not real
numbers, as well as $t_{0}$ is not a positive real number.

Figure \ref{figPdf3Dm04}\ illustrates the $3$-dimensional PDF of
$(\alpha ,H_{0},\bar{A})$, using a $3D$ density plot composed by
voxels (volume elements) with transparency and colour attributes,
specific to the $3$ free parameter case of the generalized
Chaplygin gas model with $k=0$ and $\Omega _{m0}=0.04$ (the other
five cases with $3$, $4$ and $5$ free parameters have similar $3D$
shapes). It is clearly visible the non-Gaussian behaviour of the
$3$-dimensional PDF, with the orange and green $3D$ credible
surfaces sharpening as $\alpha $ increases. Table
\ref{tableBestFitGCG} tells us that the maximum is located at
$(\alpha ,H_{0},\bar{A})=(27.0,62.7,1-5.5\times 10^{-9})$ but it
is located at a very narrow region ($H_{0}$ is extremely peaked
and $1-1\times 10^{-8}<\bar{A}<1$ with non-negligible PDF)
therefore it would be invisible even if the graphics had the range
$-1<\alpha \leqslant 28$ (because the needed resolution in the
$\bar{A}$ axis would require approximately $10^{8}$ voxels in this
dimension and only one voxel would be opaque and violet among
them). Instead, in figure \ref{figPdf3Dm04}
we see a opaque violet region indicating a maximum near $(\alpha ,H_{0},\bar{%
A})=(3.0,62,0.95)$ because there is non-negligible volume with high PDF
values. The conclusion is simple: as $\alpha $ increases the high PDF
regions narrow in width until becoming invisible, like the blue, green and
yellow credible surfaces (corresponding to constant PDF levels) do in
sequence as shown by figure \ref{figPdf3Dm04}.

\begin{figure}[!t]
\begin{minipage}[t]{0.48\linewidth}
\includegraphics[width=\linewidth]{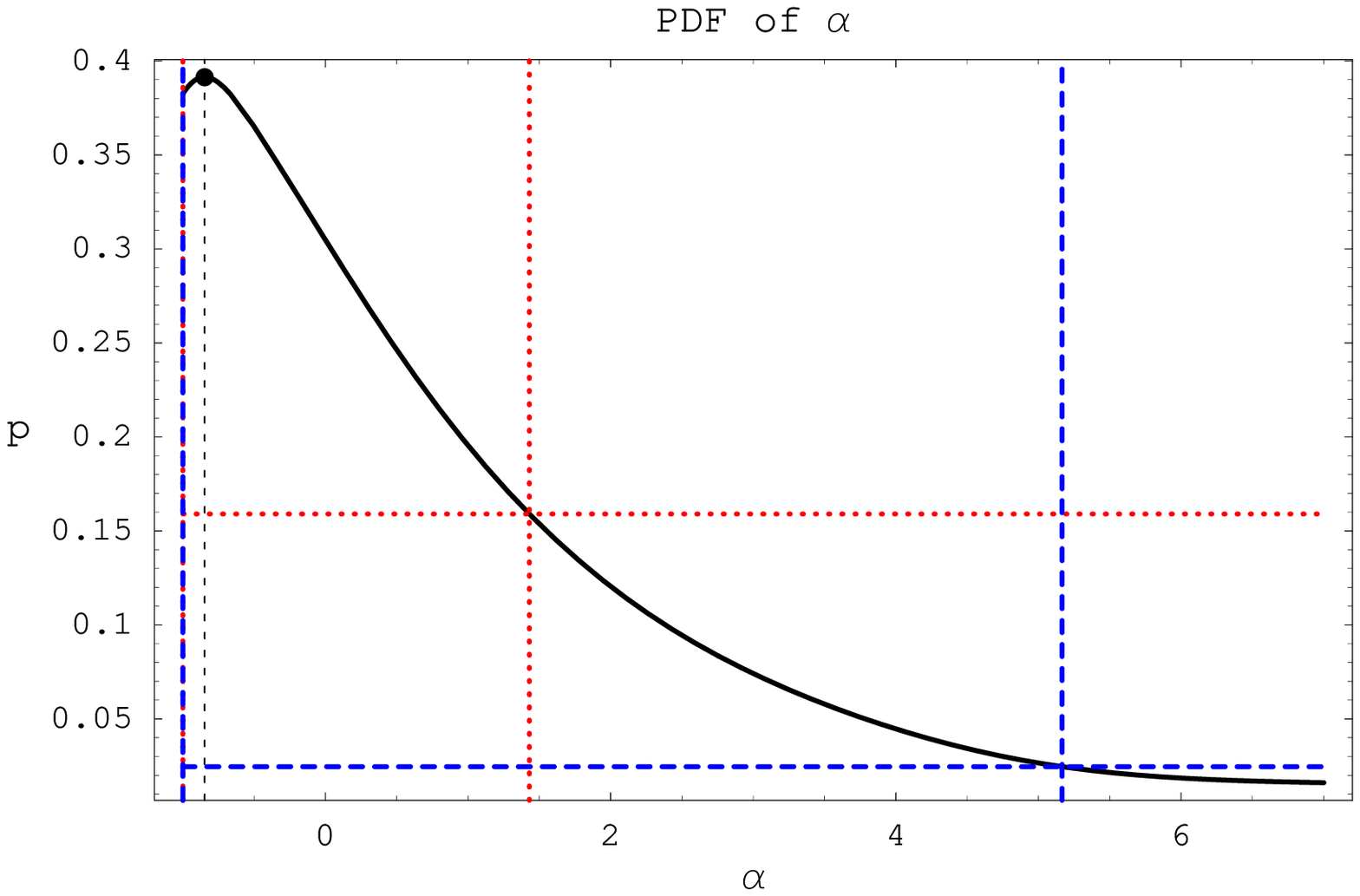}
\end{minipage} \hfill
\begin{minipage}[t]{0.48\linewidth}
\includegraphics[width=\linewidth]{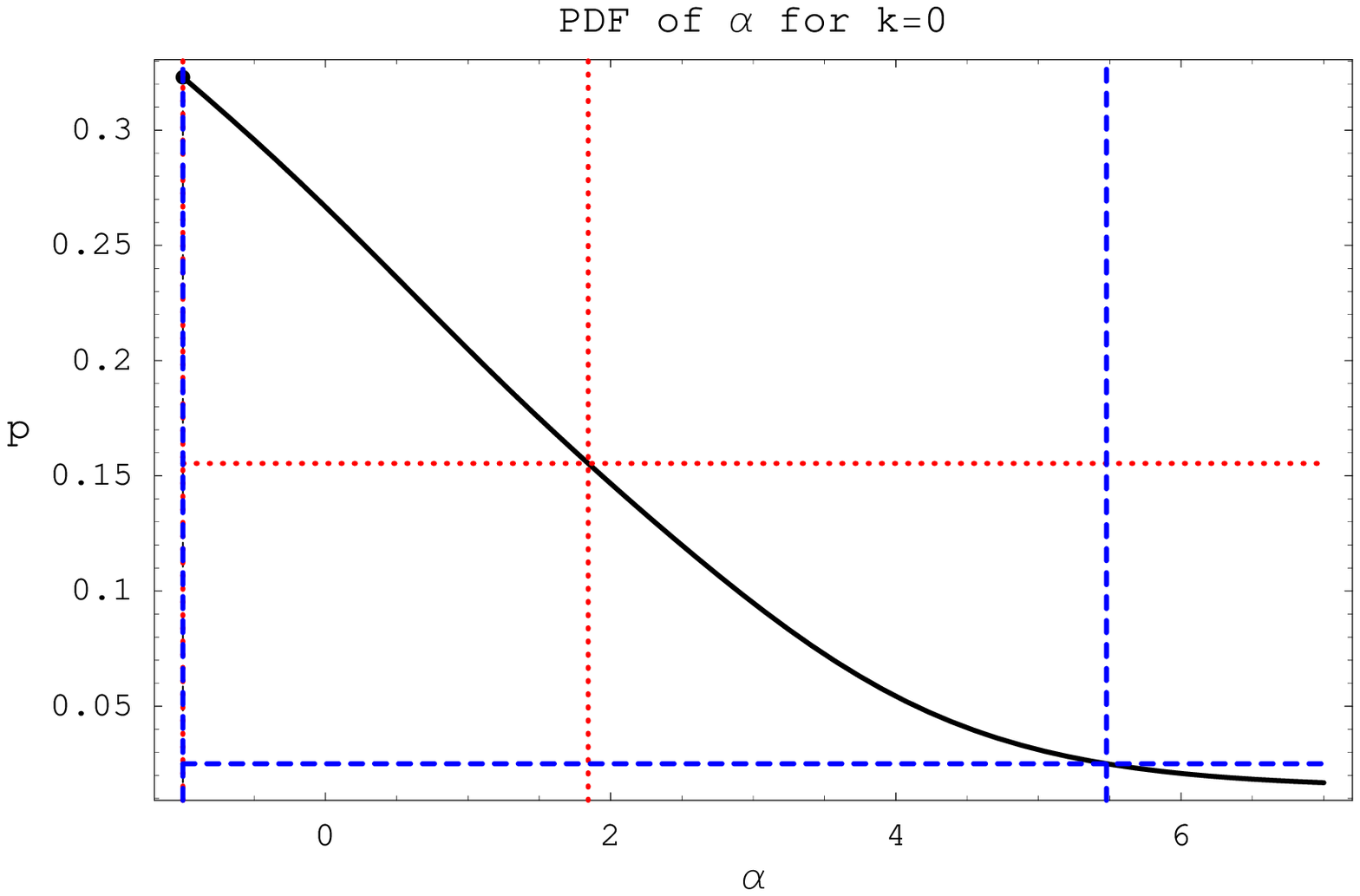}
\end{minipage} \hfill
\begin{minipage}[t]{0.48\linewidth}
\includegraphics[width=\linewidth]{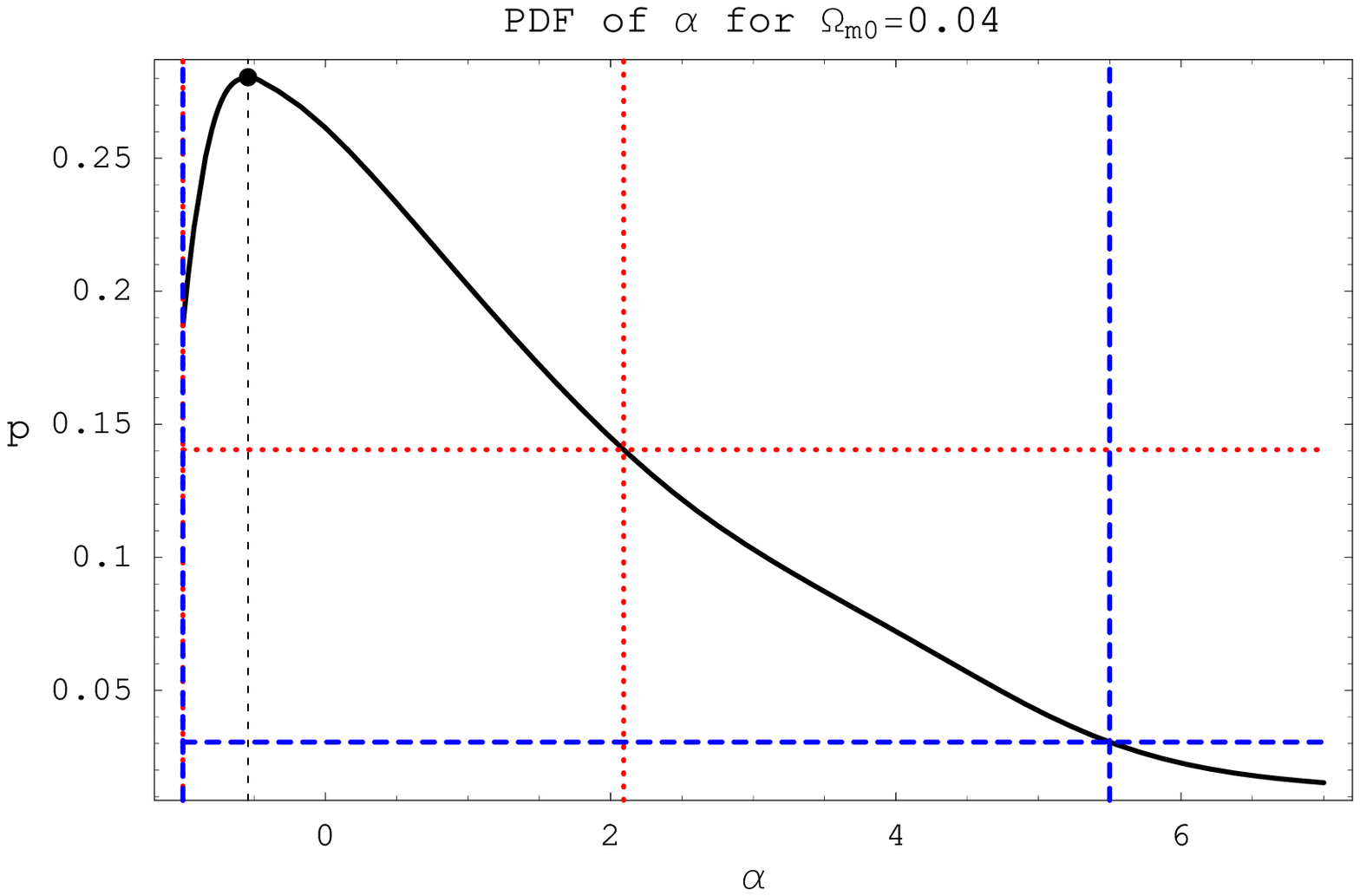}
\end{minipage} \hfill
\begin{minipage}[t]{0.48\linewidth}
\includegraphics[width=\linewidth]{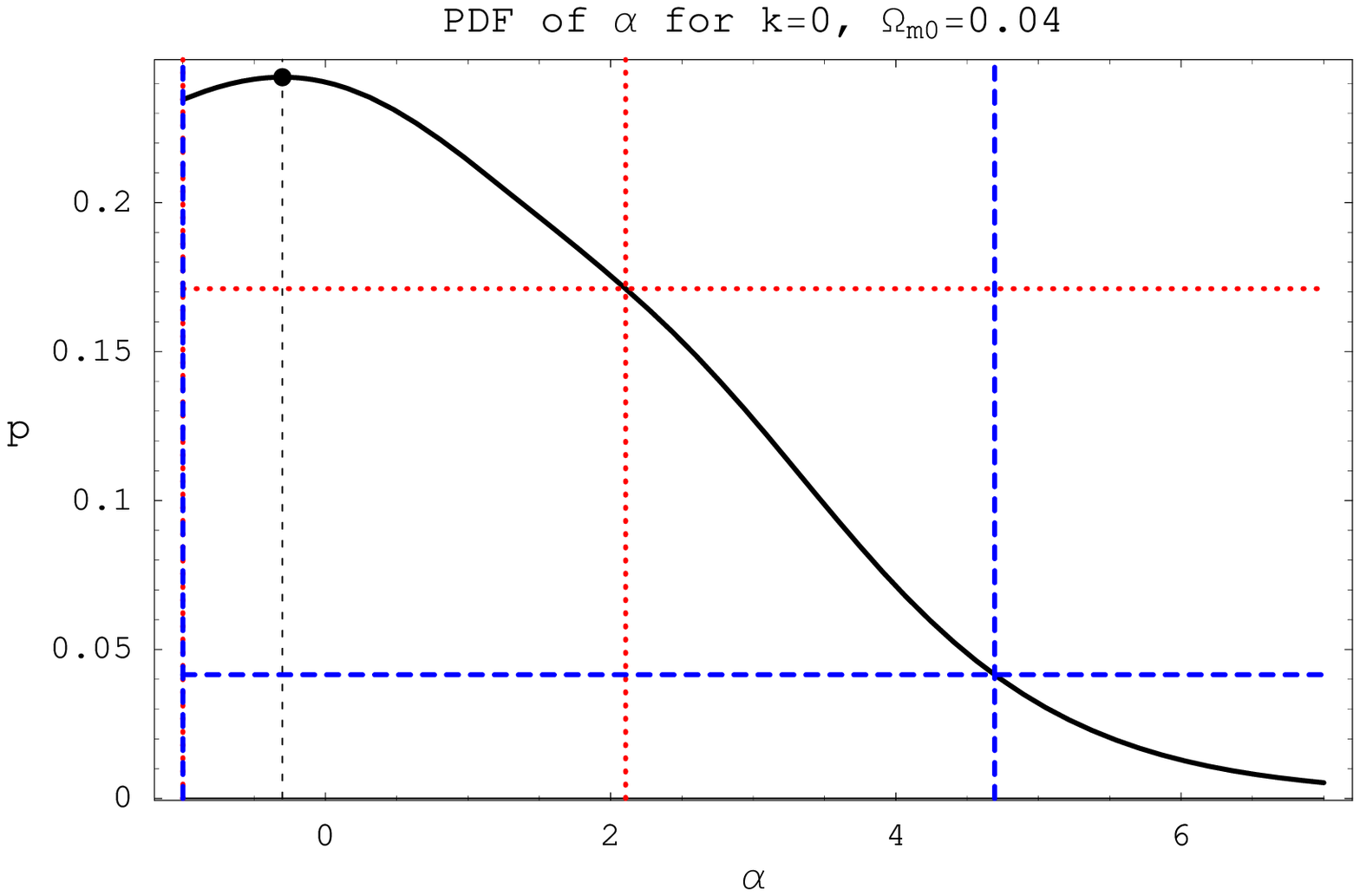}
\end{minipage} \hfill
\caption{{\protect\footnotesize The PDF of $\protect\alpha$ for the
generalized Chaplygin gas model. The solid lines are the PDF, the $1\protect%
\sigma $ ($68.27\%$) regions are delimited by red dotted lines and the $2
\protect\sigma $ ($95.45\%$) credible regions are given by blue dashed
lines. The cases for $\Omega _{m0}=0$ are not shown here because they are
similar to the ones with $\Omega _{m0}=0.04$. }}
\label{figsAlpha}
\end{figure}

The above behaviour is important to understand the results of the
marginalization processes. For example, when the marginalization is applied
to figure \ref{figPdf3Dm04}, i.e, resulting the bottom-right plot of figures
\ref{figsAlphaA}. The integration of $p(\alpha ,H_{0},\bar{A}\mid \mu _{0})$
over the $H_{0}$ parameter space yields $p(\alpha ,\bar{A}\mid \mu _{0})$,
the PDF maximum moves to lower values of $\alpha $, $(\alpha ,\bar{A}%
)=(2.75,0.95)$, because the high PDF regions for higher values of $\alpha $
have a small width with respect to $H_{0}$ and weight less than larger
regions due to the integration over $H_{0}$. Taking the marginalization
process even further, i.e., marginalizing $p(\alpha ,\bar{A}\mid \mu _{0})$
over the $\bar{A}$ space to produce $p(\alpha \mid \mu _{0})$ illustrated by
the bottom-right plot in figure \ref{figsAlpha}, the PDF maximum is once
again at lower $\alpha $, $\alpha =-0.30$, because the high PDF regions for
higher values of $\alpha $ have a small width now with respect to $\bar{A}$
and weight less than larger regions due to the integration over $\bar{A}$.

\begin{table}[t]
\begin{center}
\begin{tabular}{|c|c|c|c|c|c|c|}
\hline\hline
&  &  &  &  &  &  \\[-7pt]
& GCGM & GCGM : & GCGM : & GCGM : & GCGM : $k=0$, & GCGM : $k=0$, \\
&  & $k=0$ & $\Omega_{m0}=0$ & $\Omega_{m0}=0.04$ & $\Omega_{m0}=0$ & $%
\Omega_{m0}=0.04$ \\[2pt] \hline
&  &  &  &  &  &  \\[-7pt]
$\alpha$ & $-0.85_{-0.15}^{+6.01}$ & $-1.00_{-0.00}^{+6.48}$ & $%
-0.51_{-0.49}^{+6.39}$ & $-0.54_{-0.46}^{+6.04}$ & $0.14_{-1.14}^{+4.85}$ & $%
-0.30_{-0.70}^{+4.99}$ \\[2pt] \hline
&  &  &  &  &  &  \\[-7pt]
$H_{0}$ & $62.0_{-3.4}^{+3.3}$ & $61.3_{-2.8}^{+2.9}$ & $61.6_{-3.4}^{+3.4}$
& $61.8_{-3.5}^{+3.3}$ & $61.9_{-2.9}^{+2.7}$ & $61.8_{-2.8}^{+2.7}$ \\%
[2pt] \hline
&  &  &  &  &  &  \\[-7pt]
$\Omega_{k0}$ & $-1.26_{-1.42}^{+1.32}$ & $0$ & $0.21_{-1.68}^{+0.79}$ & $%
0.08_{-1.61}^{+0.88}$ & $0$ & $0$ \\[2pt] \hline
&  &  &  &  &  &  \\[-7pt]
$\Omega_{m0}$ & $0.00_{-0.00}^{+0.86}$ & $0.00_{-0.00}^{+0.36}$ & $0$ & $%
0.04 $ & $0$ & $0.04$ \\[2pt] \hline
&  &  &  &  &  &  \\[-7pt]
$\Omega_{c0}$ & $1.39_{-1.25}^{+1.21}$ & $1.00_{-0.36}^{+0.0}$ & $%
0.79_{-0.79}^{+1.68}$ & $0.88_{-0.88}^{+1.61}$ & $1$ & $0.96$ \\[2pt] \hline
&  &  &  &  &  &  \\[-7pt]
$\bar{A}$ & $1.00_{-0.39}^{+0.00}$ & $1.00_{-0.34}^{+0.00}$ & $%
0.94_{-0.46}^{+0.06} $ & $0.95_{-0.43}^{+0.05}$ & $0.94_{-0.35}^{+0.06}$ & $%
0.94_{-0.33}^{+0.06}$ \\[2pt] \hline
&  &  &  &  &  &  \\[-7pt]
$t_{0}$ & $15.3_{-3.2}^{+4.2}$ & $14.9_{-2.2}^{+4.8}$ & $14.3_{-2.4}^{+7.1}$
& $14.2_{-2.0}^{+6.6}$ & $14.2_{-1.8}^{+6.9}$ & $14.2_{-1.7}^{+7.5}$ \\%
[2pt] \hline
&  &  &  &  &  &  \\[-7pt]
$q_{0}$ & $-0.80_{-0.62}^{+0.86}$ & $-0.59_{-0.37}^{+0.31}$ & $%
-0.67_{-0.86}^{+0.84}$ & $-0.61_{-0.93}^{+0.78}$ & $-0.80_{-0.18}^{+0.43}$ &
$-0.89_{-0.10}^{+0.52}$ \\[2pt] \hline
&  &  &  &  &  &  \\[-7pt]
$a_{i}$ & $0.67_{-0.37}^{+0.25}$ & $0.57_{-0.25}^{+0.33}$ & $%
0.77_{-0.77}^{+0.12}$ & $0.77_{-0.54}^{+0.16}$ & $0.75_{-0.64}^{+0.14}$ & $%
0.74_{-0.49}^{+0.13}$ \\[2pt] \hline
&  &  &  &  &  &  \\[-7pt]
$p(\alpha>0)$ & $0.92\,\sigma$ & $1.05\,\sigma$ & $1.18\,\sigma$ & $%
1.12\,\sigma$ & $1.25\,\sigma$ & $1.18\,\sigma$ \\[2pt] \hline
&  &  &  &  &  &  \\[-7pt]
$p(\Omega_{k0}<0)$ & $2.20\,\sigma$ & $-$ & $0.71\,\sigma$ & $0.80\,\sigma$
& $-$ & $-$ \\[2pt] \hline
&  &  &  &  &  &  \\[-7pt]
$p(q_{0}<0)$ & $2.21\,\sigma$ & $3.31\,\sigma$ & $1.66\,\sigma$ & $%
1.75\,\sigma$ & $3.65\,\sigma$ & $3.72\,\sigma$ \\[2pt] \hline
&  &  &  &  &  &  \\[-7pt]
$p(a_{i}<1)$ & $2.22\,\sigma$ & $3.35\,\sigma$ & $1.66\,\sigma$ & $%
1.75\,\sigma$ & $3.77\,\sigma$ & $3.71\,\sigma$ \\[2pt] \hline
&  &  &  &  &  &  \\[-7pt]
$p(\dot{a}>0)$ & $5.83\,\sigma$ & $\infty\,\sigma$ & $5.88\,\sigma$ & $%
5.92\,\sigma$ & $\infty\,\sigma$ & $\infty\,\sigma$ \\[2pt] \hline\hline
\end{tabular}
\end{center}
\caption{The estimated parameters for the generalized Chaplygin
gas model (GCGM) and some specific cases of spatial section and
matter content. We use the Bayesian analysis to obtain the peak of
the one-dimensional marginal probability and the
$2\,\protect\sigma $ credible region for each parameter. $H_{0}$
is given in $km/M\!pc.s$, $\bar{A}$ in units of $c$, $t_{0}$ in
$Gy$ and $a_{i}$ in units of $a_{0}$.} \label{tableParEstGCG}
\end{table}

Table \ref{tableParEstGCG} is very comprehensive and resumes our results,
listing all the Bayesian parameter estimations for six different cases of
spatial section and matter content within the context of generalized
Chaplygin gas model. The following sub-sections will analyse in detail this
table and accompanying figures. More emphasis is spent into the analyses of
the generalized Chaplygin gas parameter $\alpha $, because it is the subject
of constant debate.

In figures \ref{figsAlpha}, \ref{figsA} and \ref{figsOmegas}, the
one-dimensional PDF shapes are matched against half-Gaussians to estimate
the probabilities outside the left and right edges, in this way increasing
the precision of the parameter estimations without excessively enlarging the
parameter space.

\subsection{$\protect\alpha$ -- the generalized Chaplygin gas parameter}

\begin{figure}[!t]
\begin{minipage}[t]{0.48\linewidth}
\includegraphics[width=\linewidth]{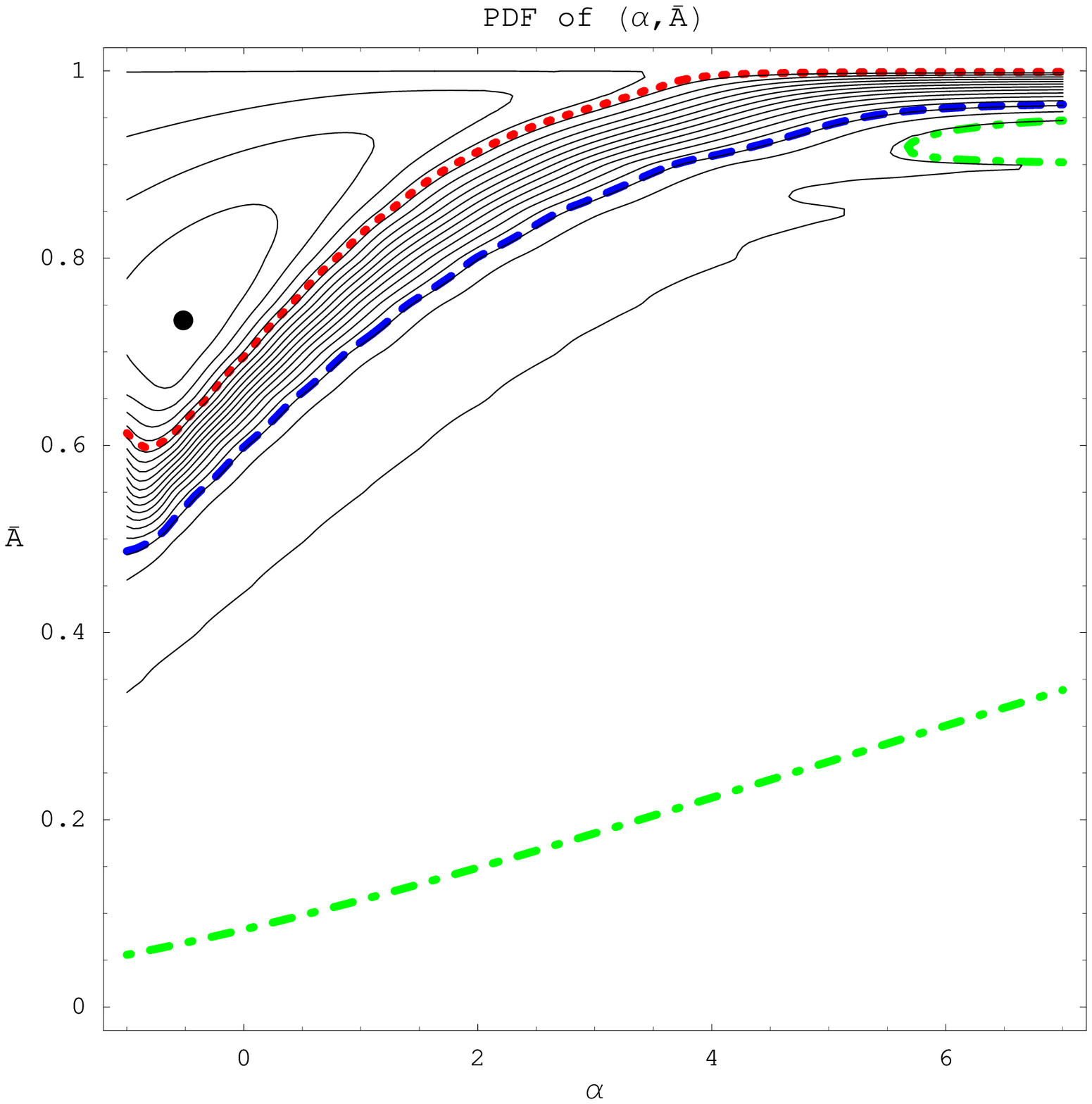}
\end{minipage} \hfill
\begin{minipage}[t]{0.48\linewidth}
\includegraphics[width=\linewidth]{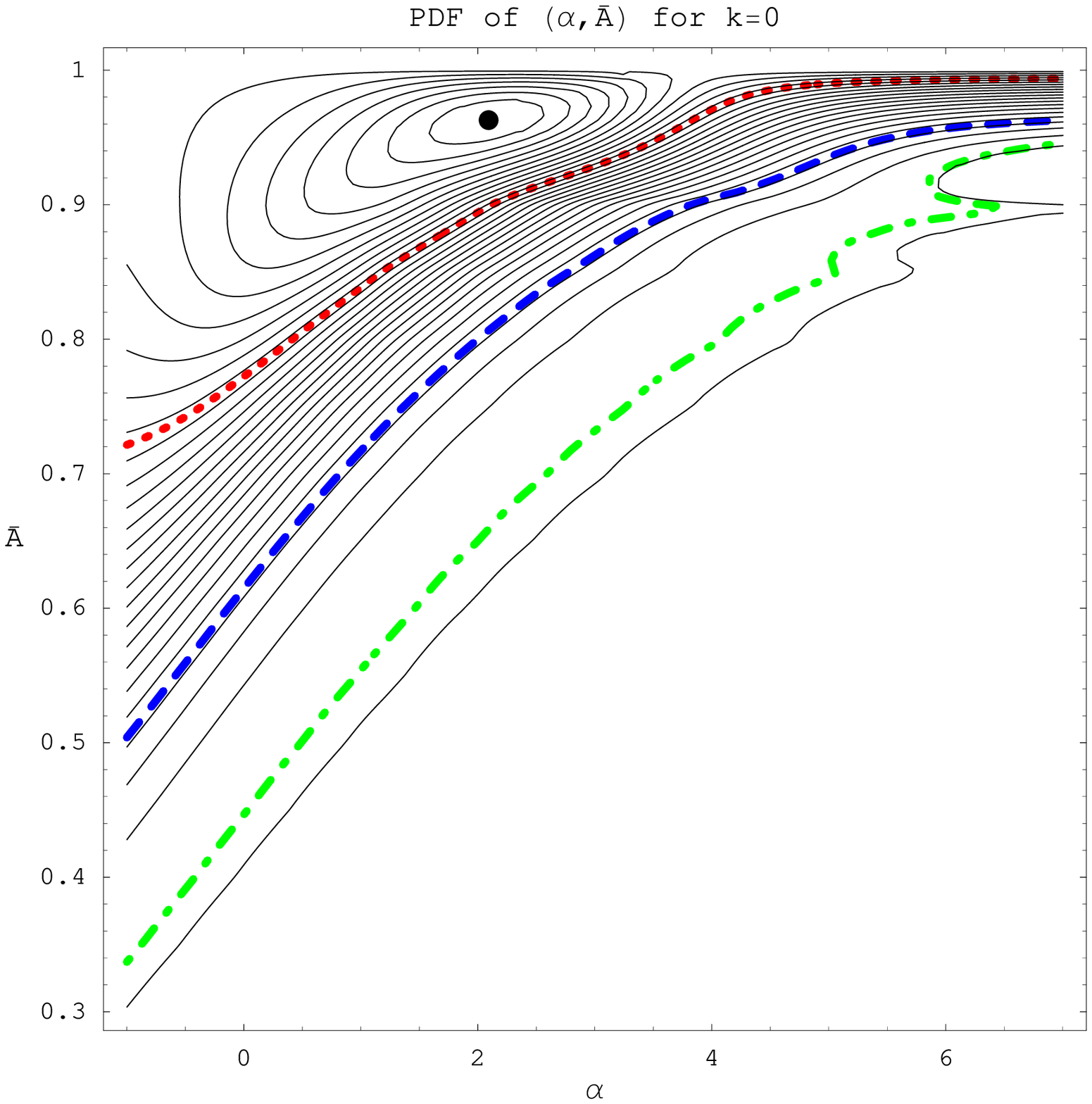}
\end{minipage} \hfill
\begin{minipage}[t]{0.48\linewidth}
\includegraphics[width=\linewidth]{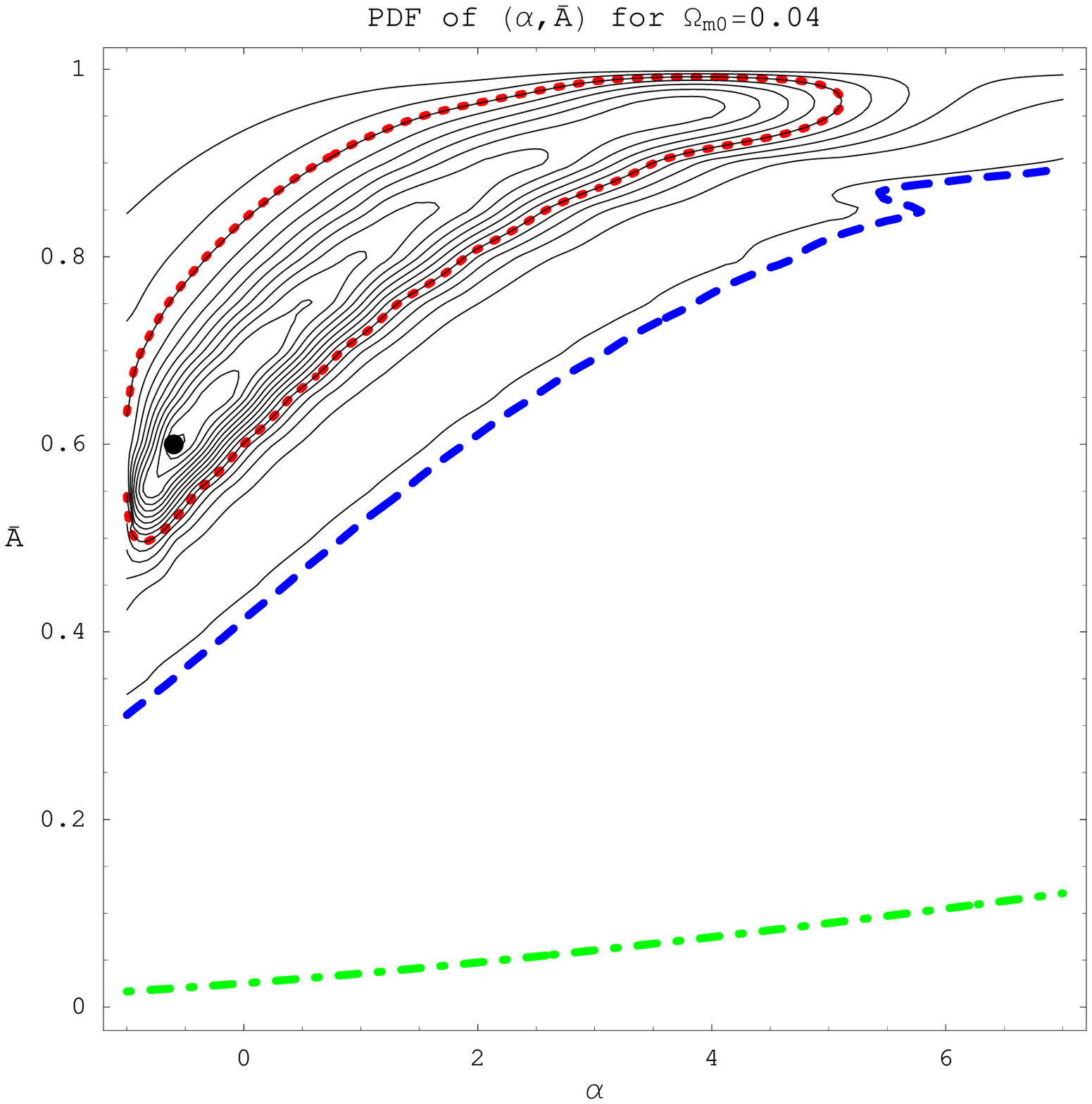}
\end{minipage} \hfill
\begin{minipage}[t]{0.48\linewidth}
\includegraphics[width=\linewidth]{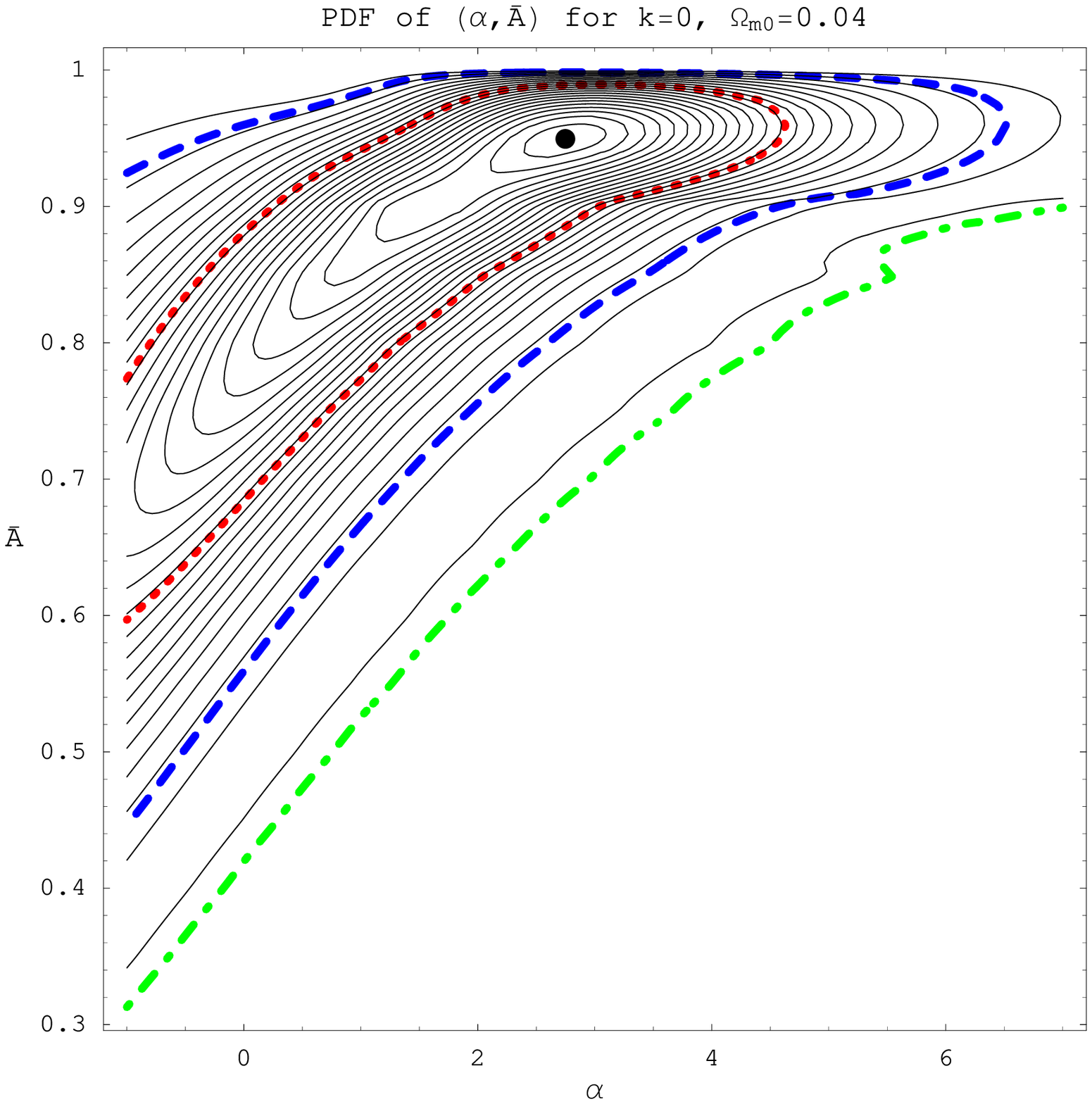}
\end{minipage} \hfill
\caption{{\protect\footnotesize The graphics of the joint PDF as function of
$(\protect\alpha,\bar{A})$ for the generalized Chaplygin gas model. The
joint PDF peak is shown by the large dot, the credible regions of $1\,%
\protect\sigma $ ($68,27\%$) by the red dotted line, the $2\,\protect\sigma $
($95,45\%$) in blue dashed line and the $3\,\protect\sigma $ ($99,73\%$) in
green dashed-dotted line. The cases for $\Omega _{m0}=0$ are not shown here
because they are similar to the ones with $\Omega _{m0}=0.04$. }}
\label{figsAlphaA}
\end{figure}

The parameter estimation of $\alpha $ is strongly related to the $\bar{A}$
parameter, i.e., the joint PDFs for the parameter space of $(\alpha ,\bar{A}%
) $ do not have a Gaussian shape, so the marginalization process changes the
peak values and credible regions. Table \ref{tableParEstGCG}, figures \ref
{figsAlphaA} and \ref{figsAlpha} are carefully analysed below with respect
to $\alpha $. In brief, the estimations for $\alpha $ are quite spread
(i.e., they have large credible regions) and they differ a lot depending on
if two or one-dimensional parameter space is used.

\subsubsection{Generalized Chaplygin gas model with $5$ free parameters}

In figures \ref{figsAlphaA}, the GCGM with $5$ free parameters has a joint
PDF peak of $0.948$ at $(\alpha ,\bar{A})=(-0.52,0.73)$ with $1\,\sigma $ ($%
68,27\,\%$), $2\,\sigma $ ($95,45\,\%$) and $3\sigma $ ($99,73\,\%$)
credible region contours with PDF values of $0.717$, $0.123$ and $0.002$,
respectively. The two-dimensional credible regions are not Gaussian-shaped,
and their extremes give the estimations $(\alpha ,\bar{A}%
)=(-0.52_{-0.48}^{>+7.52},0.73_{-0.13}^{+0.27})$ at $1\,\sigma $, $(\alpha ,%
\bar{A})=(-0.52_{-0.48}^{>+7.52},0.73_{-0.24}^{+0.27})$ at $2\,\sigma $, and
$(\alpha ,\bar{A})=(-0.52_{-0.48}^{>+7.52},0.73_{-0.67}^{+0.27})$ at $%
3\,\sigma $ confidence levels. The symbol ($>$) in the upper
credible value for $\alpha $ is due to the fact that the graphics
is limited to $-1<\alpha \leqslant 7$.

The marginalization (integration) of $p(\alpha ,\bar{A}\mid \mu _{0})$ over
the $\bar{A}$ space yields $p(\alpha \mid \mu _{0})$, shown in figure \ref
{figsAlpha}, where the credible intervals are $\alpha =-0.85_{-0.15}^{+2.28}$
at $1\,\sigma $ and $\alpha =-0.85_{-0.15}^{+6.01}$ at $2\,\sigma $ (the $%
3\,\sigma $ PDF level cannot be seen with $\alpha \leqslant 7$). The PDF
peak of $0.383$ at $\alpha =-0.85$ is greater than the PDF value of $0.195$
for $\alpha =1$ (ordinary Chaplygin gas model), so the region with PDF
greater than $0.195$ has a CDF of $0.607$, i.e., we can say that $\alpha =1$
is ruled out at a $60.7\,\%$ ($0.85\,\sigma $)\ confidence level (or it is
preferred at $39.3\,\%$). Table \ref{tableParEstGCG} shows that positive
values of $\alpha $ are more likely at a $64.1\,\%$ ($0.92\,\sigma $)\
confidence level.

\subsubsection{GCGM with $4$ free parameters and $k=0$}

The $4$-parameter case with $k=0$ has a joint PDF peak of $1.165$ at $%
(\alpha ,\bar{A})=(2.09,0.96)$ with $1\,\sigma $, $2\,\sigma $ and $3\sigma $
credible region contour levels of $0.695$, $0.141$ and $0.007$,
respectively. The two-dimensional credible regions give the estimations $%
(\alpha ,\bar{A})=(2.09_{-3.09}^{>+4.91},0.96_{-0.24}^{+0.04})$ at $%
1\,\sigma $, $(\alpha ,\bar{A})=(2.09_{-3.09}^{>+4.91},0.96_{-0.44}^{+0.04})$
at $2\,\sigma $, and $(\alpha ,\bar{A}%
)=(2.09_{-3.09}^{>+4.91},0.96_{-0.62}^{+0.04})$ at $3\,\sigma $ confidence
levels.

The estimation based on $p(\alpha \mid \mu _{0})$ yields $\alpha
=-1.00_{-0.00}^{+2.84}$ at $1\,\sigma $ and $\alpha =-1.00_{-0.00}^{+6.48}$
at $2\,\sigma $ confidence levels. The PDF peak of $0.323$ at $\alpha =-1.00$
is greater than the PDF value of $0.205$ for $\alpha =1$ such that $\alpha
=1 $ is ruled out at a $53.1\,\%$ ($0.72\,\sigma $)\ confidence level.
Positive values of $\alpha $ are preferred at a $70.5\,\%$ ($1.05\,\sigma $%
)\ confidence level.

\subsubsection{GCGM with $4$ free parameters and $\Omega _{m0}=0$}

The joint PDF has a peak value of $1.239$ at $(\alpha ,\bar{A})=(-0.51,0.60)$
with $1\,\sigma $, $2\,\sigma $ and $3\sigma $ credible region contour
levels of $0.378$, $0.017$ and $0.007$, respectively. These levels indicate
that the joint two-dimensional PDF is strongly peaked. The two-dimensional
credible regions give the estimations $(\alpha ,\bar{A}%
)=(-0.51_{-0.45}^{+4.97},0.60_{-0.12}^{+0.40})$ at $1\,\sigma $, $(\alpha ,%
\bar{A})=(-0.51_{-0.49}^{>+7.51},0.60_{-0.35}^{+0.40})$ at $2\,\sigma $, and
$(\alpha ,\bar{A})=(-0.51_{-0.49}^{>+7.51},0.60_{-0.58}^{+0.40})$ at $%
3\,\sigma $ confidence levels.

The estimations based on one-dimensional PDF yield the same peak value (due
to the highly peaked joint PDF), with $\alpha =-0.51_{-0.46}^{+2.86}$ at $%
1\,\sigma $ and $\alpha =-0.51_{-0.49}^{+6.39}$ at $2\,\sigma $ confidence
levels. The PDF peak of $0.263$ at $\alpha =-0.51$ is greater than the PDF
value of $0.200$ for $\alpha =1$, and we can say that $\alpha =1$ is
preferred at a $55.9\,\%$ ($0.77\,\sigma $)\ confidence level. Positive
values of $\alpha $ are preferred at a $76.2\,\%$ ($1.18\,\sigma $)\
confidence level.

\subsubsection{GCGM with $4$ free parameters and $\Omega _{m0}=0.04$}

The joint PDF has a peak value of $1.257$ at $(\alpha ,\bar{A})=(-0.56,0.60)$
with $1\,\sigma $, $2\,\sigma $ and $3\sigma $ credible region contour
levels of $0.428$, $0.020$ and $0.005$, respectively, which show a highly
peaked behaviour. The two-dimensional credible regions give the estimations $%
(\alpha ,\bar{A})=(-0.56_{-0.44}^{+5.69},0.60_{-0.10}^{+0.40})$ at $%
1\,\sigma $, $(\alpha ,\bar{A}%
)=(-0.56_{-0.44}^{>+7.56},0.60_{-0.29}^{+0.40}) $ at $2\,\sigma $, and $%
(\alpha ,\bar{A})=(-0.56_{-0.44}^{>+7.56},0.60_{-0.58}^{+0.40})$ at $%
3\,\sigma $ confidence levels.

The estimations based on one-dimensional PDF give $\alpha
=-0.54_{-0.46}^{+2.63}$ at $1\,\sigma $ and $\alpha =-0.54_{-0.46}^{+6.04}$
at $2\,\sigma $ confidence levels. The PDF peak of $0.280$ at $\alpha =-0.54$
is greater than the PDF value of $0.202$ for $\alpha =1$, and we can say
that $\alpha =1$ is preferred at a $50.9\,\%$ ($0.69\,\sigma $)\ confidence
level. Positive values of $\alpha $ are preferred at a $73.6\,\%$ ($%
1.12\,\sigma $)\ confidence level.

\subsubsection{GCGM with $3$ free parameters and $k=0$, $\Omega _{m0}=0$}

The joint PDF has a peak value of $1.336$ at $(\alpha ,\bar{A})=(3.11,0.95)$
with $1\,\sigma $, $2\,\sigma $ and $3\sigma $ credible region contour
levels of $0.583$, $0.115$ and $0.007$, respectively. The two-dimensional
credible regions give the estimations $(\alpha ,\bar{A}%
)=(3.11_{-4.11}^{+1.86},0.95_{-0.38}^{+0.04})$ at $1\,\sigma $, $(\alpha ,%
\bar{A})=(3.11_{-4.11}^{+3.72},0.95_{-0.52}^{+0.05})$ at $2\,\sigma $, and $%
(\alpha ,\bar{A})=(3.11_{-4.11}^{>+3.89},0.95_{-0.65}^{+0.05})$ at $%
3\,\sigma $ confidence levels.

The estimations based on one-dimensional PDF give $\alpha
=0.14_{-1.14}^{+2.22}$ at $1\,\sigma $ and $\alpha =0.14_{-1.14}^{+4.85}$ at
$2\,\sigma $ confidence levels. The PDF peak of $0.222$ at $\alpha =0.14$ is
greater than the PDF value of $0.208$ for $\alpha =1$, and we can say that $%
\alpha =1$ is preferred at a $63.2\,\%$ ($0.90\,\sigma $)\ confidence level.
Positive values of $\alpha $ are preferred at a $78.8\,\%$ ($1.25\,\sigma $%
)\ confidence level.

\subsubsection{GCGM with $3$ free parameters and $k=0$, $\Omega _{m0}=0.04$}

The joint PDF has a peak value of $1.367$ at $(\alpha ,\bar{A})=(2.75,0.95)$
with $1\,\sigma $, $2\,\sigma $ and $3\sigma $ credible region contour
levels of $0.595$, $0.113$ and $0.007$, respectively. The two-dimensional
credible regions give the estimations $(\alpha ,\bar{A}%
)=(2.75_{-4.75}^{+1.88},0.95_{-0.35}^{+0.04})$ at $1\,\sigma $, $(\alpha ,%
\bar{A})=(2.75_{-4.75}^{+3.77},0.95_{-0.50}^{+0.05})$ at $2\,\sigma $, and $%
(\alpha ,\bar{A})=(2.75_{-4.75}^{>+4.25},0.95_{-0.63}^{+0.05})$ at $%
3\,\sigma $ confidence levels.

The estimations based on one-dimensional PDF give $\alpha
=-0.30_{-0.70}^{+2.41}$ at $1\,\sigma $ and $\alpha =-0.30_{-0.70}^{+4.99}$
at $2\,\sigma $ confidence levels. The PDF peak of $0.242$ at $\alpha =-0.30$
is greater than the PDF value of $0.214$ for $\alpha =1$, and we can say
that $\alpha =1$ is preferred at a $53.1\,\%$ ($0.72\,\sigma $)\ confidence
level. Positive values of $\alpha $ are preferred at a $76.0\,\%$ ($%
1.18\,\sigma $)\ confidence level.

\vspace{0.1in}

The non-Gaussian shape of the credible region contours in the $(\alpha ,\bar{%
A})$ two-dimensional parameter space is the culprit of this difference for
the estimations of $\alpha $, so it is crucial to emphasize how the $\alpha $
parameter is estimated : from a joint two-dimensional PDF or an
one-dimensional PDF (after marginalization of the joint two-dimension PDF).
The most robust parameter estimation is based on the one-dimensional PDF,
therefore giving small importance to regions in the joint two-dimensions
which have a small joint PDF volume (CDF). Comparing the estimations, the
ones based on two-dimensional parameter space always give larger values for $%
\alpha $, after marginalization the PDF is peaked at lower values of $\alpha
$.

We also expect that a large sample of SNe Ia will narrow the credible
regions of $\alpha $, so the issue of a ordinary Chaplygin gas model (CGM)
being preferred or ruled out will be probably better addressed.

\subsection{The parameter $\bar{A}$}

The generalized Chaplygin gas model\ (GCGM) estimations for $\bar{A}$ are
given by table \ref{tableParEstGCG}, which if compared with the results of
the CGM (table $5$ of ref. \cite{roberto}) clearly indicate that the PDF
peak is closer to $\bar{A}=1$ and the dispersion is increased when the GCGM
is considered. This behaviour is due to the shape of the joint PDF in the
parameter space $(\alpha ,\bar{A})$, figures \ref{figsAlphaA} explicitly
show that as $\alpha $ increases there is a narrower region of high PDF
values closer to higher values of $\bar{A}$.

\begin{figure}[!t]
\begin{minipage}[t]{0.48\linewidth}
\includegraphics[width=\linewidth]{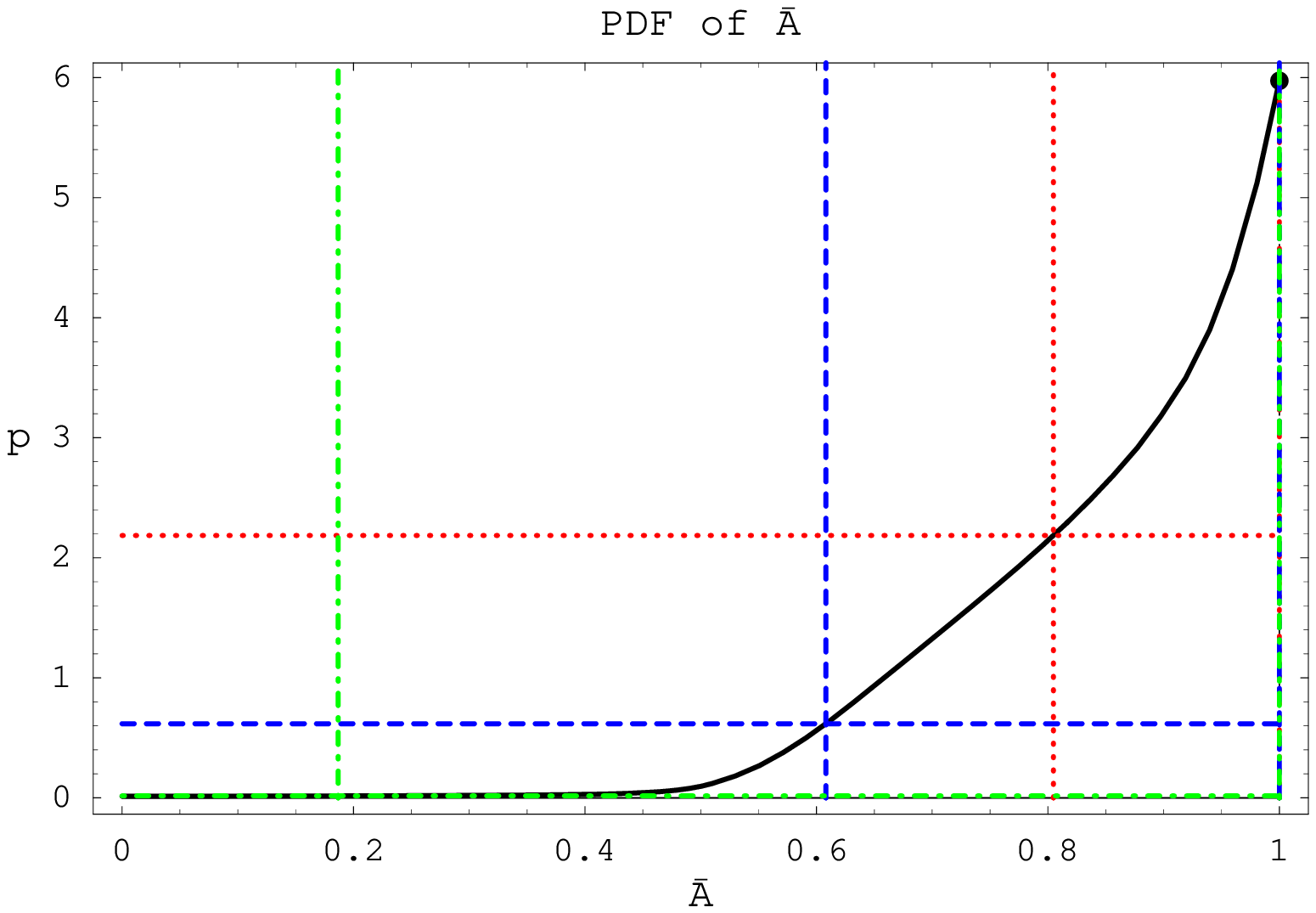}
\end{minipage} \hfill
\begin{minipage}[t]{0.48\linewidth}
\includegraphics[width=\linewidth]{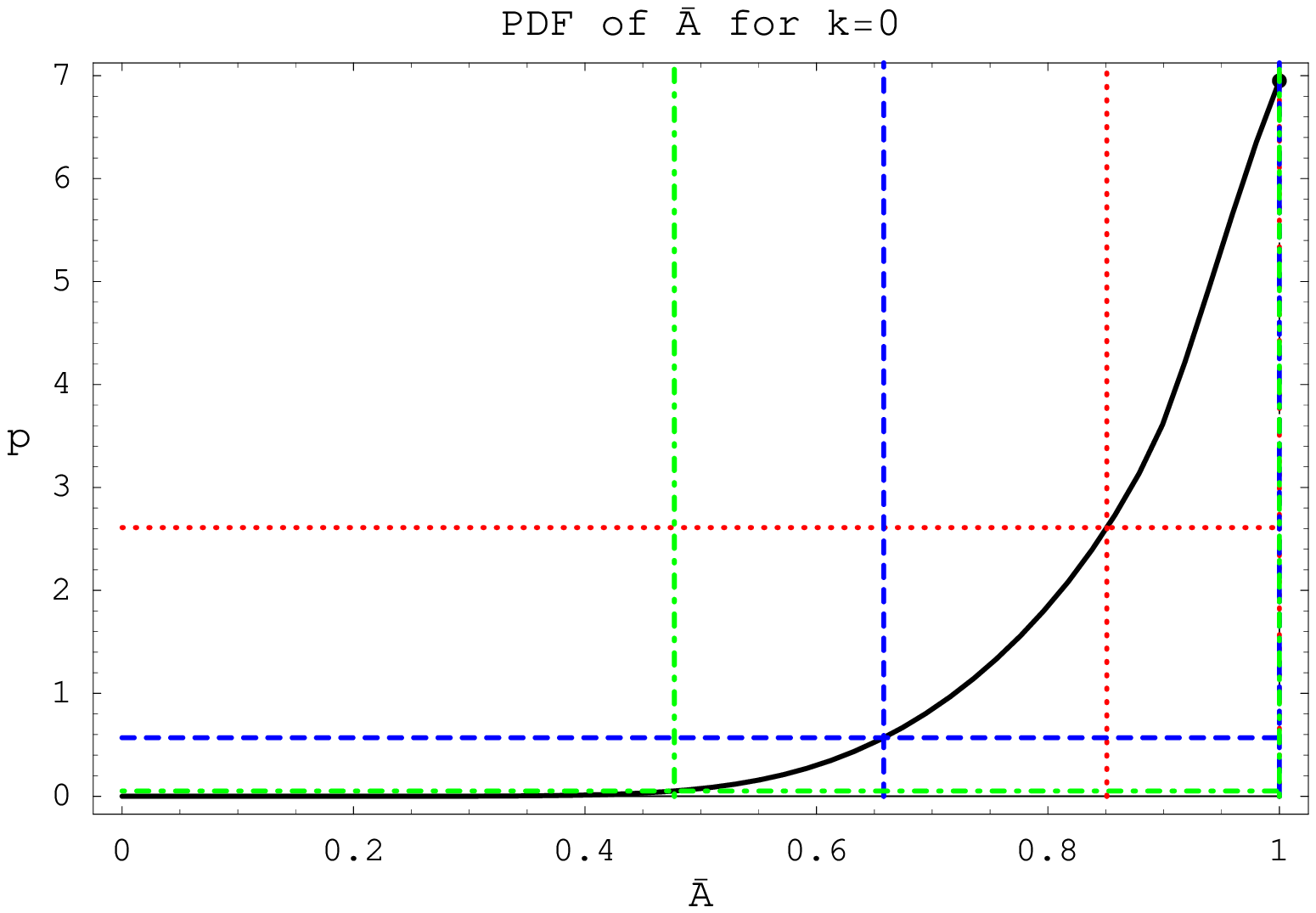}
\end{minipage} \hfill
\begin{minipage}[t]{0.48\linewidth}
\includegraphics[width=\linewidth]{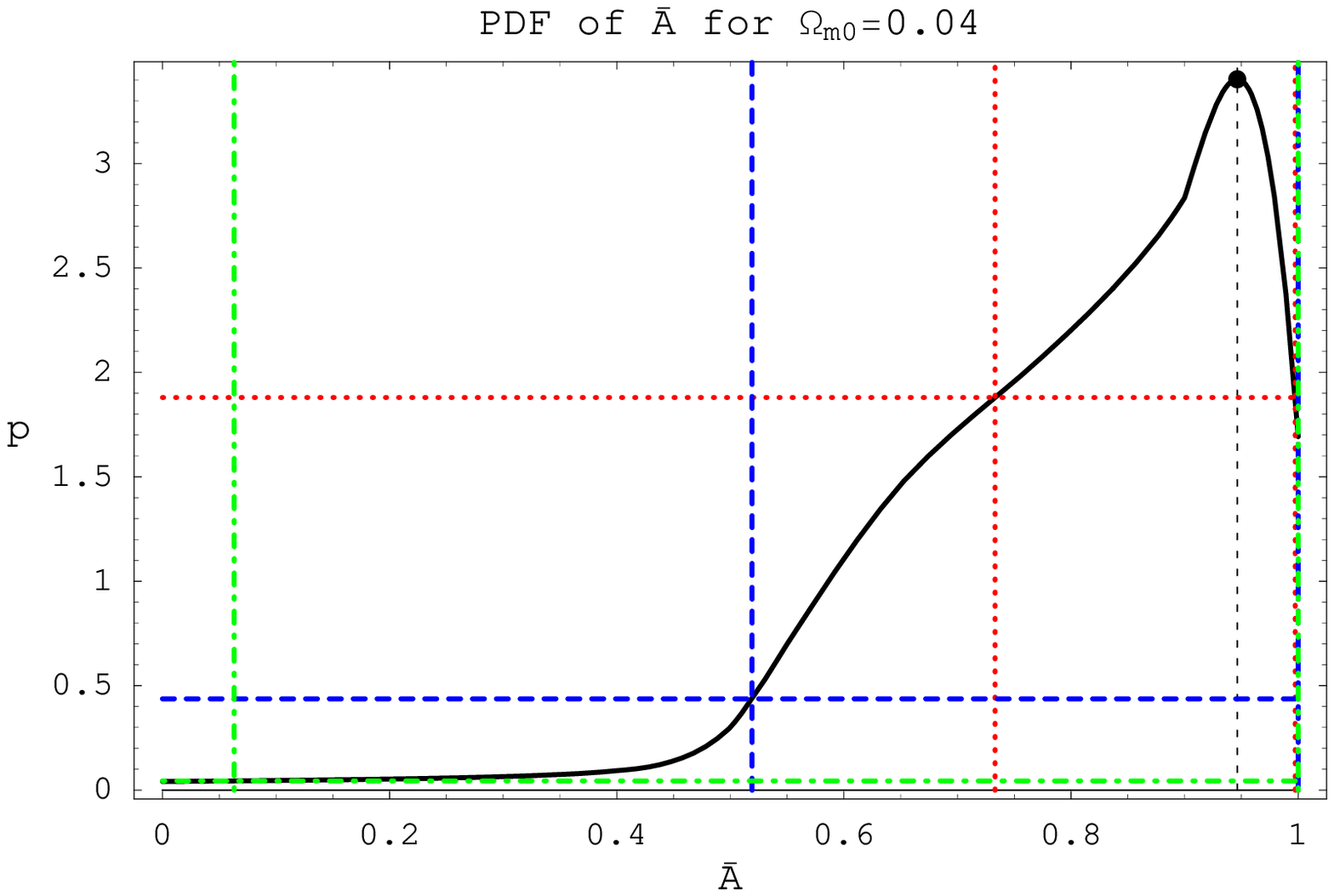}
\end{minipage} \hfill
\begin{minipage}[t]{0.48\linewidth}
\includegraphics[width=\linewidth]{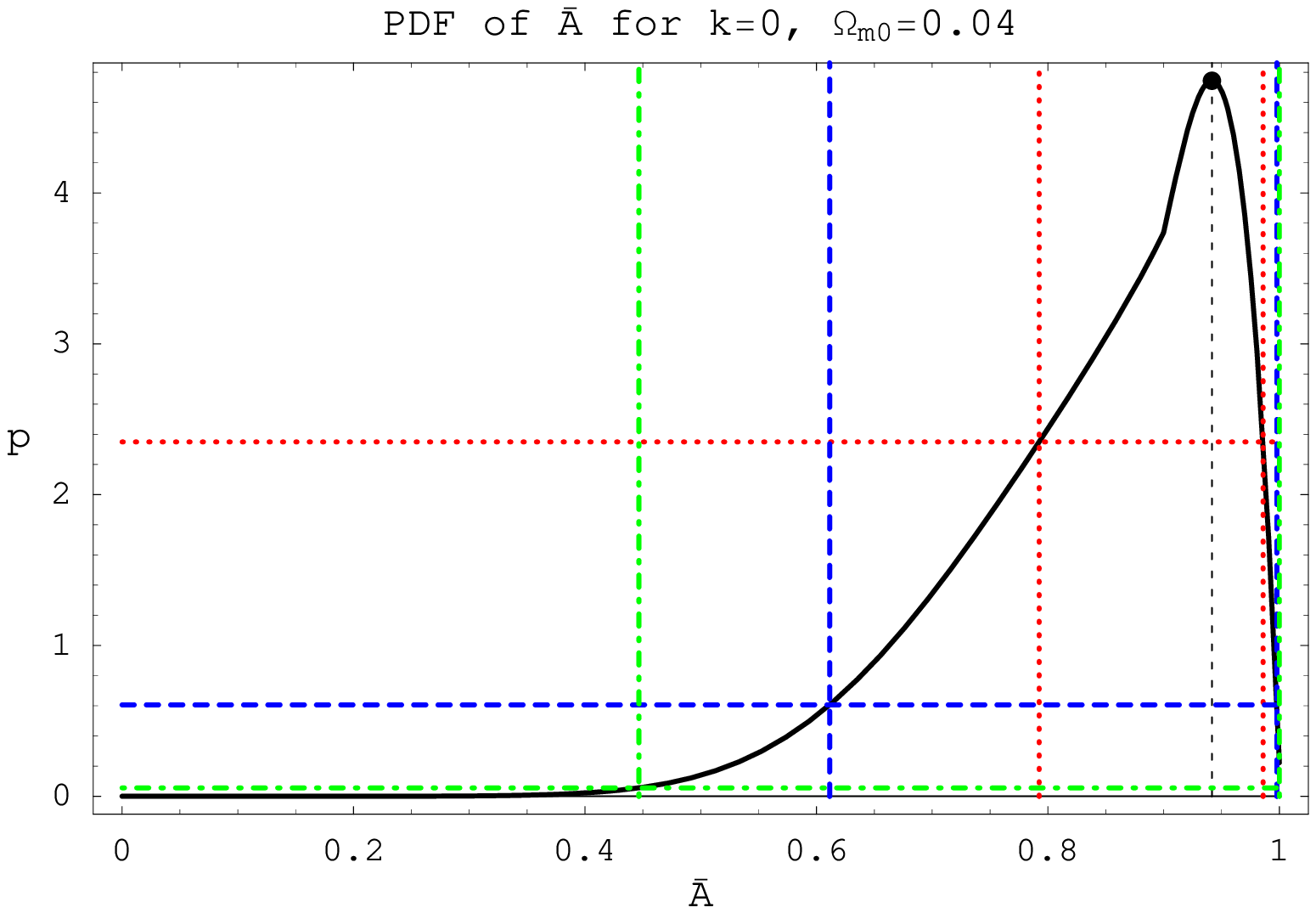}
\end{minipage} \hfill
\caption{{\protect\footnotesize The one-dimensional PDF of $\bar{A}$ for the
generalized Chaplygin gas model. The solid lines are the PDF, the $1\protect%
\sigma $ ($68.27\%$) regions are delimited by red dotted lines, the $2%
\protect\sigma $ ($95.45\%$) credible regions are given by blue dashed lines
and the $3\protect\sigma $ ($99.73\%$) regions are delimited by green
dashed-dotted lines. The cases for $\Omega _{m0}=0$ are not shown here
because they are similar to the ones with $\Omega _{m0}=0.04$. }}
\label{figsA}
\end{figure}

But the estimations for $\bar{A}$ shown in table \ref{tableParEstGCG} and
figures \ref{figsA}\ have to be interpreted. They do not indicate that the $%
\Lambda CDM$ limit ($\bar{A}=1$) is favoured because the resolution for the $%
\bar{A}$ sampling is $0.05$, i.e, a PDF\ peak at $\bar{A}=1.00$ means the
peak happens for $0.95<\bar{A}\leqslant 1$ indeed. Therefore, in figures \ref
{figsA}, the graphics for the PDF of $\bar{A}$ which are peaked at $\bar{A}%
=1 $ have the shape of the graphics for the case $\Omega _{m0}=0.04$, with
the PDF peak closer to $\bar{A}=1$. See ref. \cite{roberto} for a detailed
comparison between the CGM and $\Lambda CDM$ cases.

The analyses of $\alpha $ presented above have considered the joint PDFs for
the two-dimensional parameter space of $(\alpha ,\bar{A})$, and the extremes
of the credible regions have given estimations for $(\alpha ,\bar{A})$ at $%
1\,\sigma $, $2\,\sigma $, and $3\,\sigma $ confidence levels. These
estimations for $\bar{A}$ are almost always different from the estimations
based on the one-dimensional PDF of $\bar{A}$, obtained by marginalizing $%
p(\alpha ,\bar{A}\mid \mu _{0})$ over the $\alpha $ space. We stress again
that the most robust estimations are based on the one-dimensional PDFs. The
general behaviour of table \ref{tableParEstGCG} is that the PDF\ peak is at $%
0.95<\bar{A}\leqslant 1$ and the $2\,\sigma $ credible interval begins at $%
\bar{A}\simeq 0.5$ and finishes at $\bar{A}\simeq 1$.

\subsection{The Hubble parameter $H_{0}$}

The parameter estimation of $H_{0}$ shown in table \ref{tableParEstGCG} is
almost the same for all cases, around $62\ km/M\!pc.s$ with a narrow
credible region of $\pm 3\ km/M\!\!pc.s$. This estimation, like the one for
the CGM in table $5$ of ref. \cite{roberto}, is compatible with other SNe Ia
analyses.

\subsection{The curvature density parameter $\Omega _{k0}$}

\begin{figure}[!h]
\begin{minipage}[t]{0.48\linewidth}
\includegraphics[width=\linewidth]{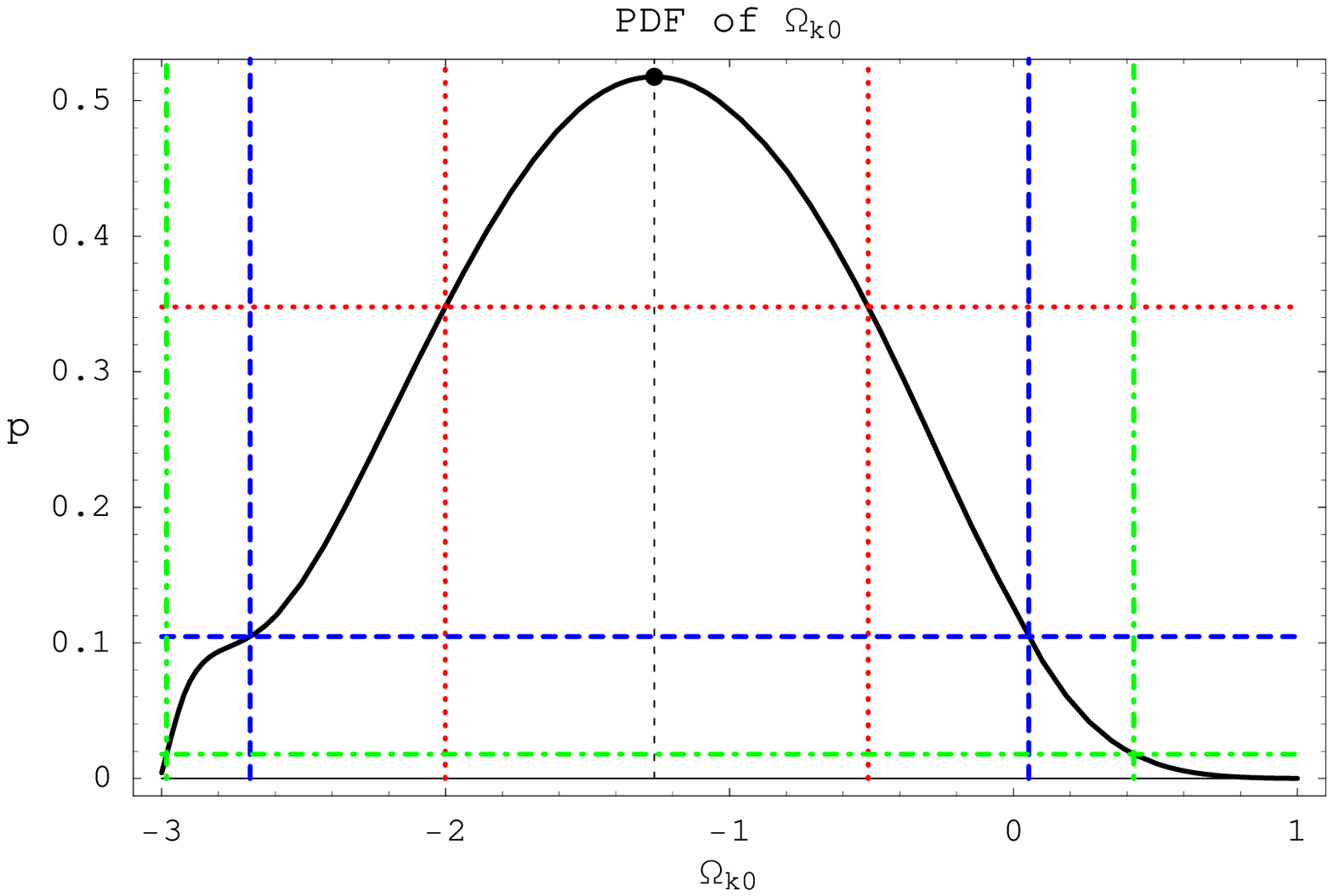}
\end{minipage} \hfill
\begin{minipage}[t]{0.48\linewidth}
\includegraphics[width=\linewidth]{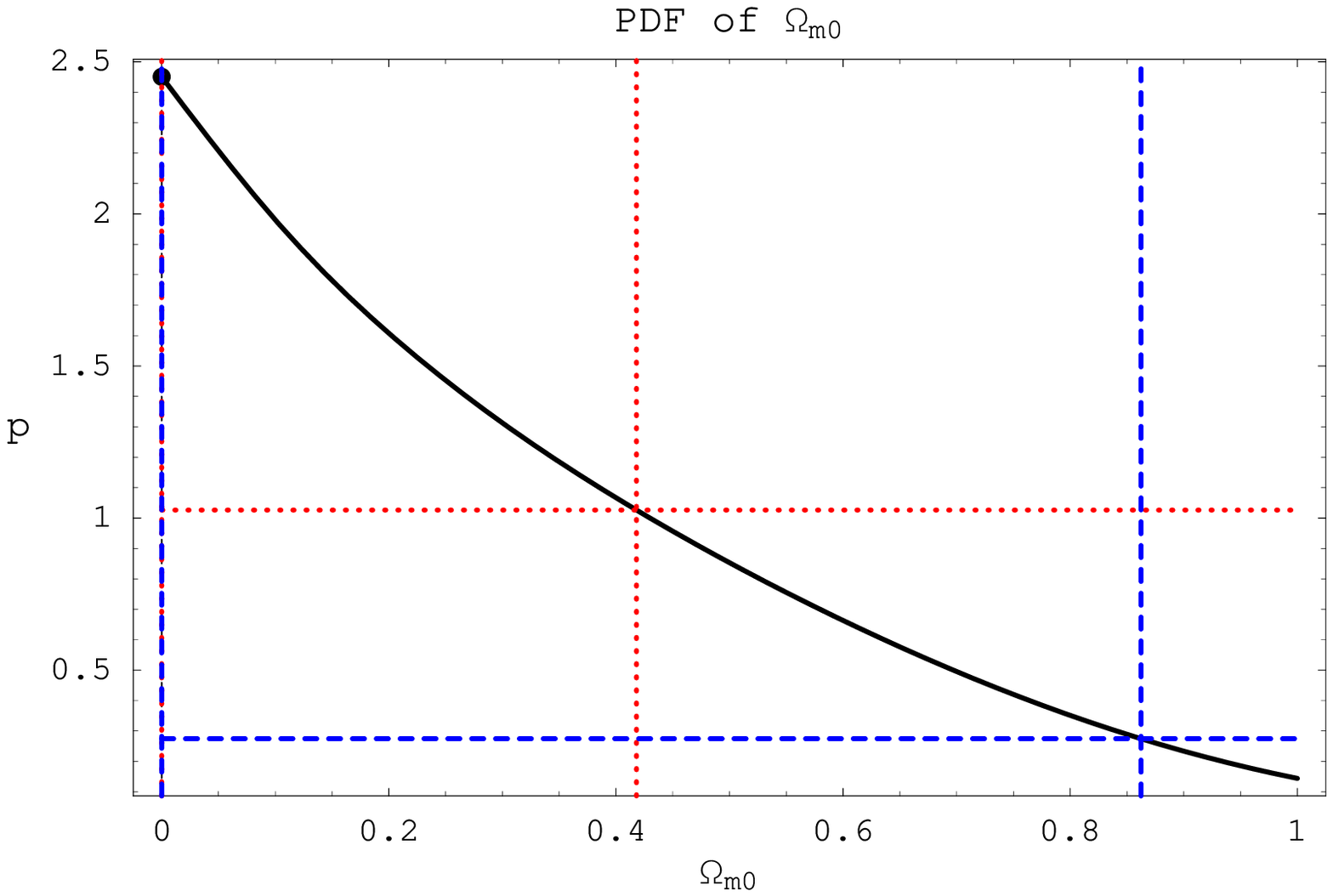}
\end{minipage} \hfill
\begin{minipage}[t]{0.48\linewidth}
\includegraphics[width=\linewidth]{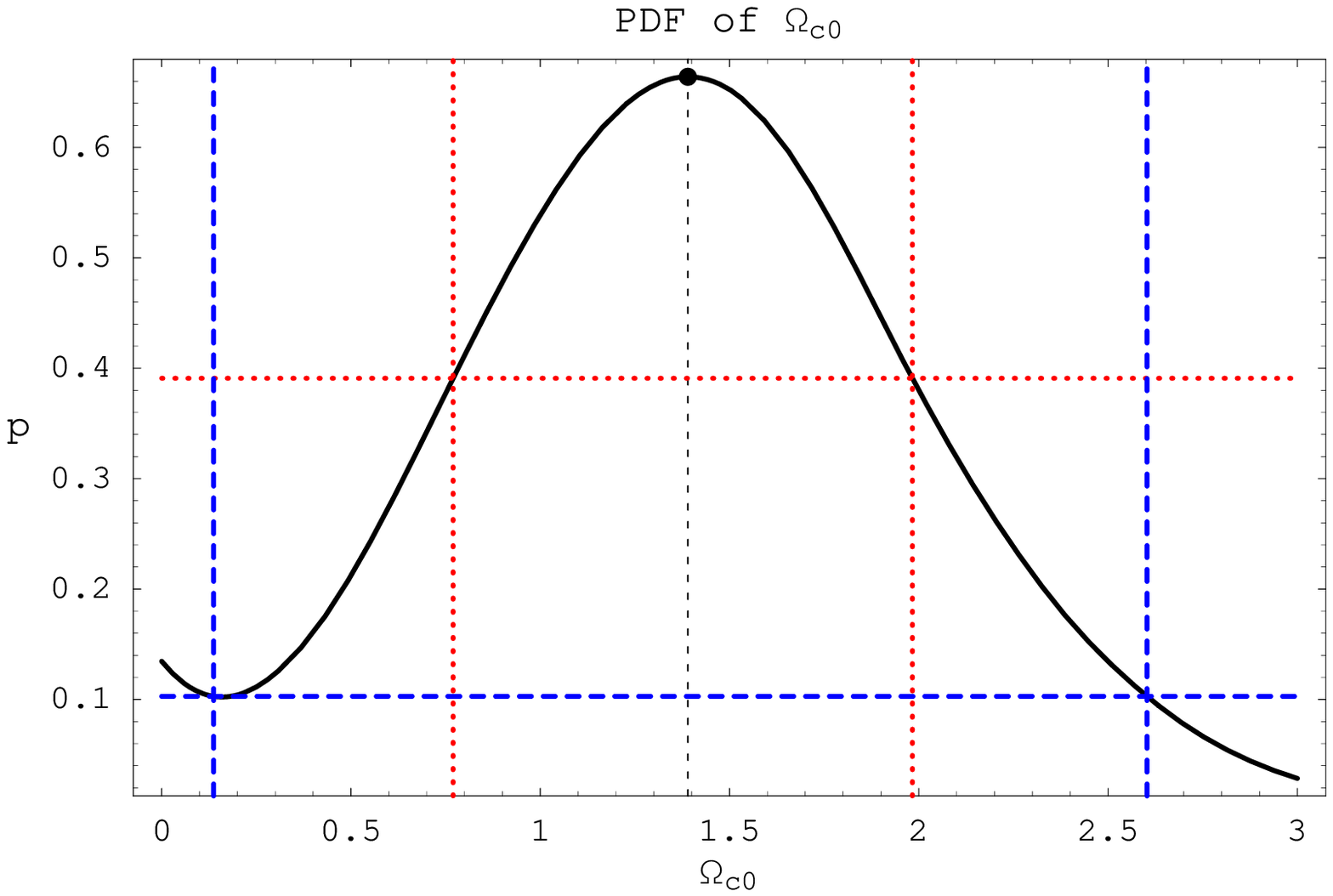}
\end{minipage} \hfill
\begin{minipage}[t]{0.48\linewidth}
\includegraphics[width=\linewidth]{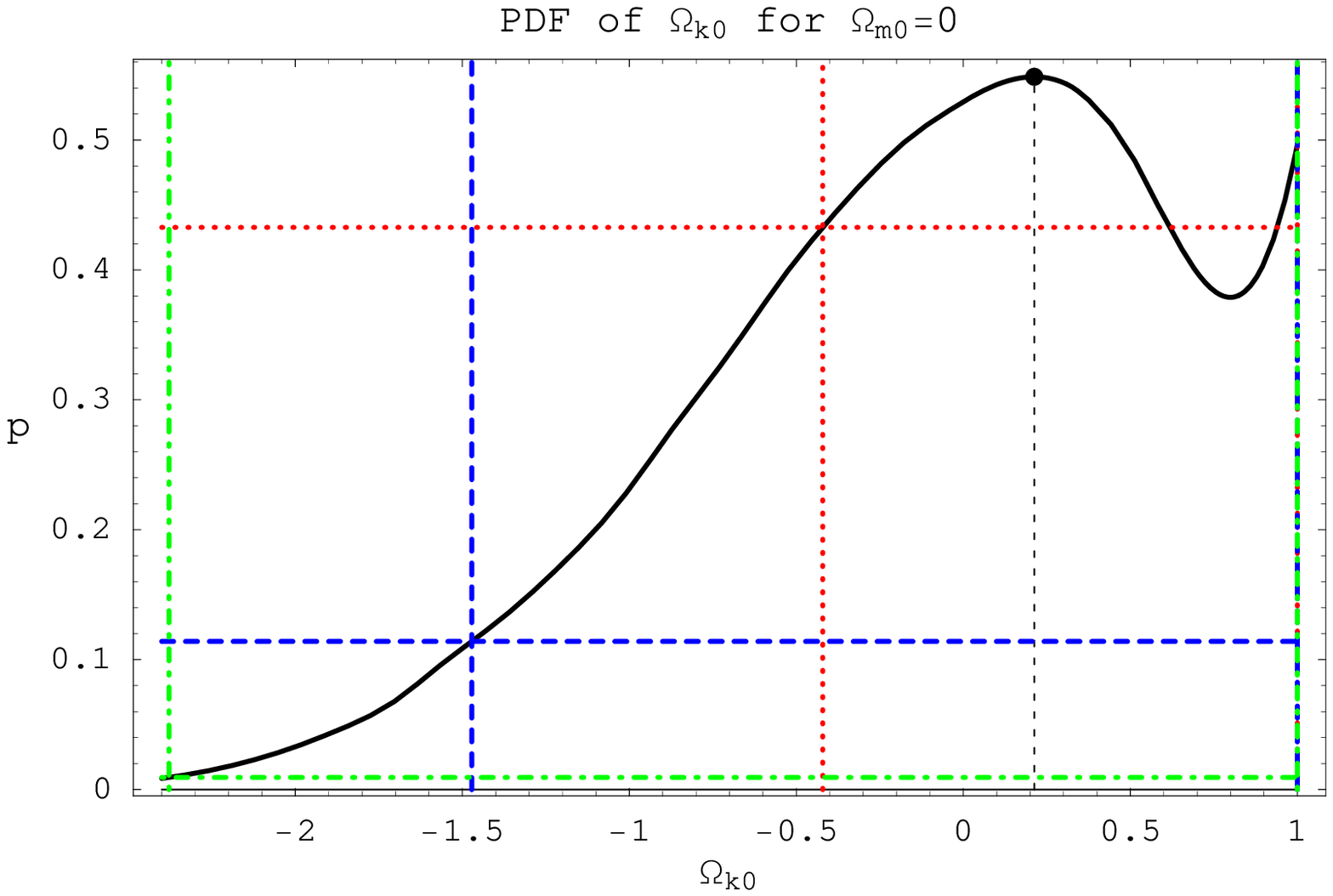}
\end{minipage} \hfill
\begin{minipage}[t]{0.48\linewidth}
\includegraphics[width=\linewidth]{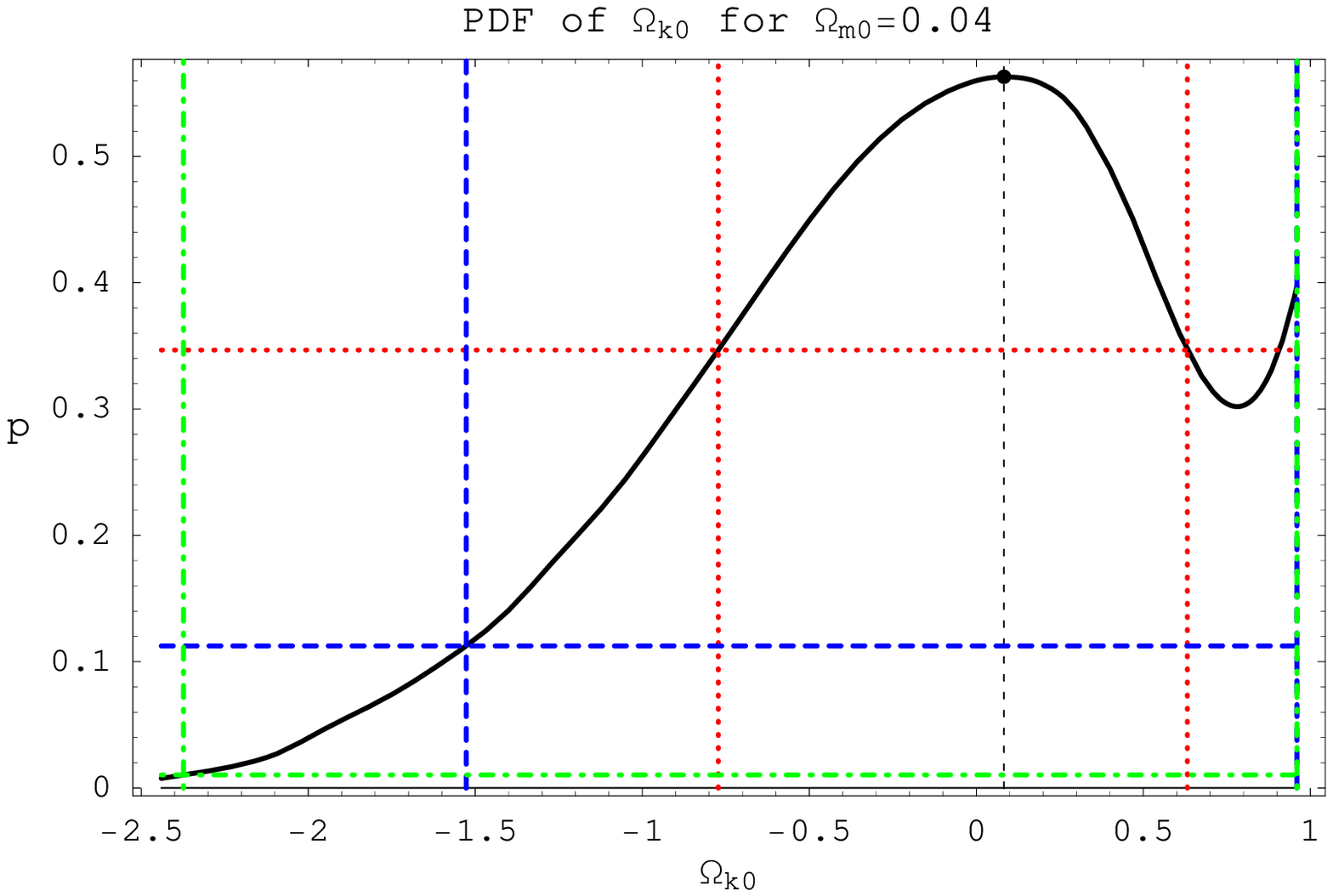}
\end{minipage} \hfill
\begin{minipage}[t]{0.48\linewidth}
\includegraphics[width=\linewidth]{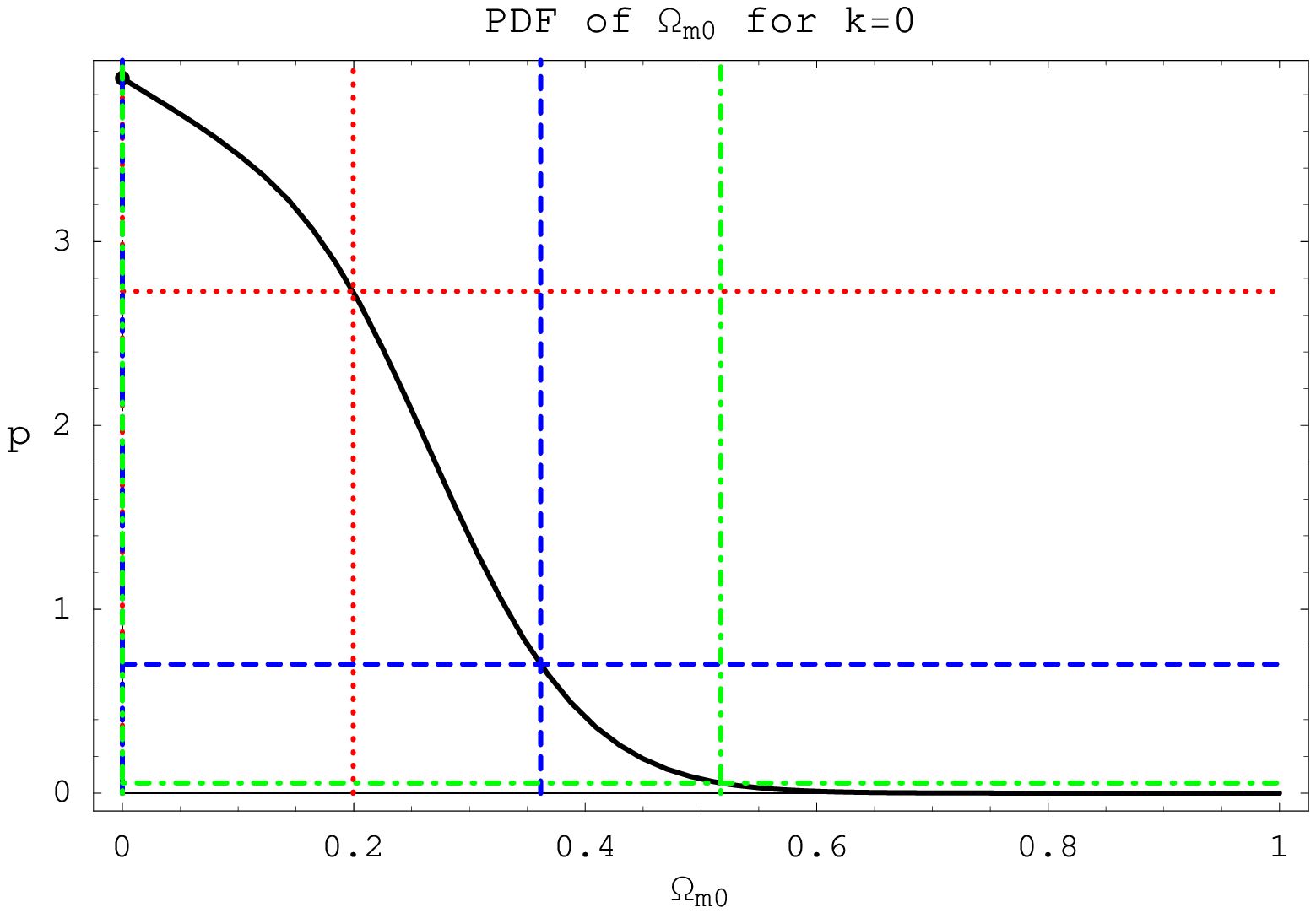}
\end{minipage} \hfill
\caption{{\protect\footnotesize The one-dimensional PDF of $\Omega _{k0}$, $%
\Omega _{m0}$ and $\Omega _{c0}$ for the generalized Chaplygin gas model.
The solid lines are the PDF, the $1\protect\sigma $ ($68.27\%$) regions are
delimited by red dotted lines, the $2\protect\sigma $ ($95.45\%$) credible
regions are given by blue dashed lines and the $3\protect\sigma $ ($99.73\%$%
) regions are delimited by green dashed-dotted lines. As $\Omega
_{c0}=1-\Omega _{k0}-\Omega _{k0}$, for $\Omega _{m0}=0$ we have $\Omega
_{c0}=1-\Omega _{k0}$, for $\Omega _{m0}=0.04$ then $\Omega
_{c0}=0.96-\Omega _{k0}$ and for $\Omega _{k0}=0$ we also have $\Omega
_{c0}=1-\Omega _{m0}$. }}
\label{figsOmegas}
\end{figure}

Table \ref{tableParEstGCG} and figure \ref{figsOmegas} show that the
curvature density parameter $\Omega _{k0}$ estimation for the GCGM is almost
like for the CGM of ref. \cite{roberto}, i.e., it also strongly depends on
the cold dark matter content of the GCGM.

\begin{figure}[!t]
\begin{center}
\includegraphics[scale=0.8]{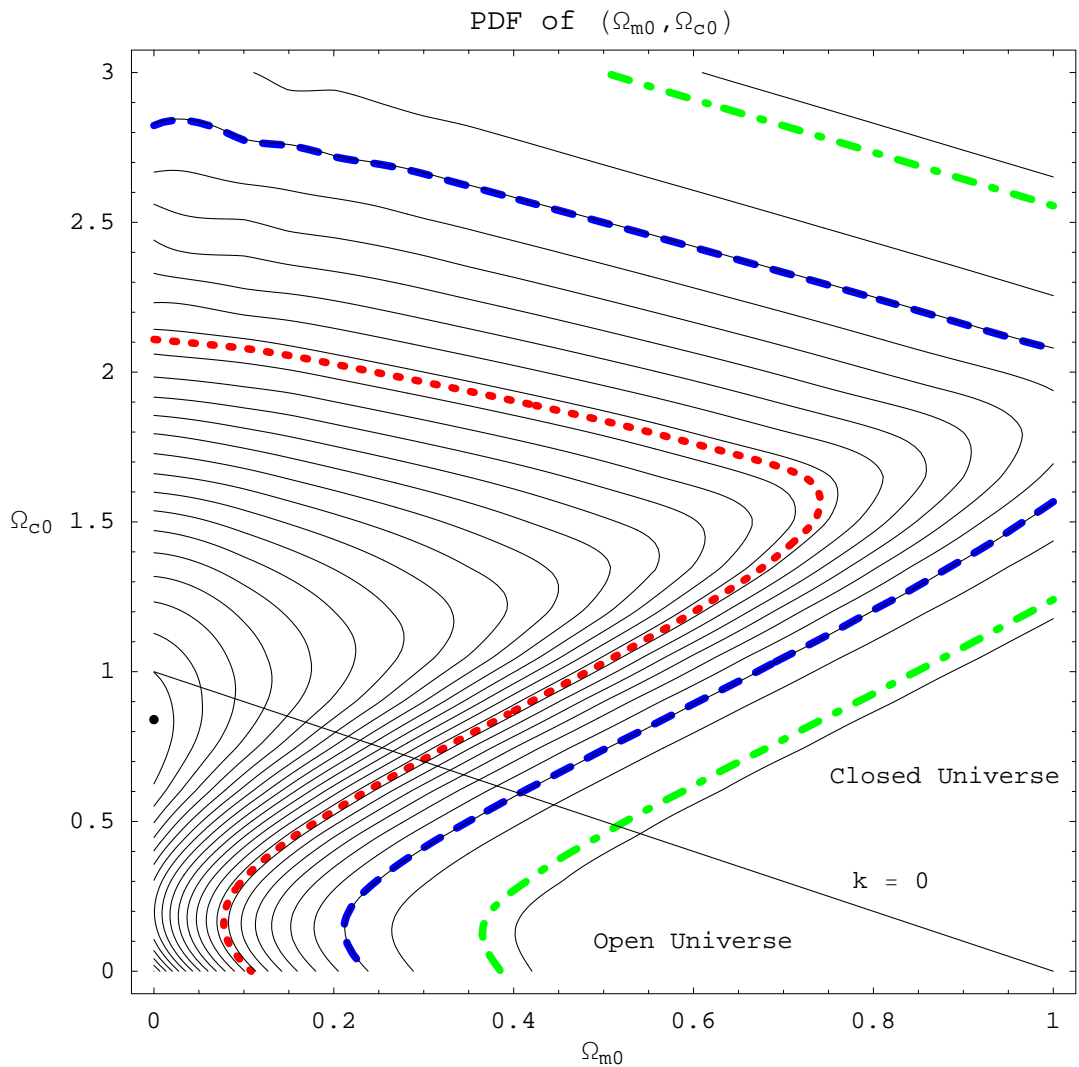}
\end{center}
\caption{{\protect\footnotesize The graphics of the joint PDF as function of
$(\Omega _{m0},\Omega _{c0})$ for the generalized Chaplygin gas model, where
$p(\Omega _{m0},\Omega _{c0})$ is a integral of $p(\protect\alpha
,H_{0},\Omega _{m0},\Omega _{c0},\bar{A})$ over the $(\protect\alpha ,H_{0},%
\bar{A})$ parameter space. It shows a different shape for the confidence
regions with respect to the $\Lambda CDM$ but similar to the CGM of ref.
\protect\cite{roberto}. The joint PDF peak has the value $1.39$ for $(\Omega
_{m0},\Omega _{c0})=(0.00,0.81)$ (shown by the large dot), the credible
regions of $1\,\protect\sigma $ ($68,27\%$, shown in red dotted line), $2\,%
\protect\sigma $ ($95,45\%$, in blue dashed line) and $3\,\protect\sigma $ ($%
99,73\%$, in green dashed-dotted line) have PDF levels of $0.50$, $0.13$\
and $0.02$, respectively. As $\Omega _{k0}+\Omega _{m0}+\Omega _{c0}=1$, the
probability for a spatially flat Universe is on the line $\Omega
_{m0}+\Omega _{c0}=1$, above it we have the region for a closed Universe ($%
k>0$, $\Omega _{k0}<0$), and below, the region for an open Universe ($k<0$, $%
\Omega _{k0}>0$). }}
\label{figOmegam0c0}
\end{figure}

In the $5$-parameter case a closed Universe ($\Omega _{k0}<0$) is preferred
at $97.19\,\%$ ($2.20\,\sigma $) and a spatially flat Universe is ruled out
at $93.44\,\%$ ($1.84\,\sigma $) confidence level, i.e., the PDF of $\Omega
_{k0}=0$ is smaller than the PDF of $-2.57<\Omega _{k0}<0$ and this region
has a CDF of $93.44\,\%$ ($1.84\,\sigma $). We can also see the figure \ref
{figOmegam0c0} that the integral of the joint PDF over the upper region
gives the same high likelihood of a closed Universe, although the joint PDF
peak is located in the open Universe region.

But assuming the hypothesis that $\Omega _{m0}=0.04$, a $4$-parameter case,
the behaviour is opposite, a closed Universe ($\Omega _{k0}<0$) is preferred
at $57.75\,\%$ ($0.80\,\sigma $) and a spatially flat Universe is preferred
at $90.32\,\%$ ($1.66\,\sigma $) confidence level, i.e., the PDF of $\Omega
_{k0}=0$ is smaller than the PDF of $0<\Omega _{k0}<0.17$ and this region
has a small CDF of $9.68\,\%$. Therefore, if we impose a low value for $%
\Omega _{m0}$ then the spatially flat Universe is favoured.

\subsection{The cold dark matter density parameter $\Omega _{m0}$}

The generalized Chaplygin gas model favours small values of $\Omega _{m0}$
like the ordinary CGM, see table \ref{tableParEstGCG} and table $5$ of ref.
\cite{roberto}. The graphics for $\Omega _{m0}$ in figure \ref{figsOmegas},
for the $5$-parameter case and $4$-parameter case with $k=0$, shows a
estimation for $\Omega _{m0}$ peaked at $0$ at a high level of confidence,
with $\Omega _{m0}<0.42$ (for the $5$-parameter case) and $\Omega _{m0}<0.20$
(for $4$-parameter case with $k=0$) at $1\,\sigma $ ($68.27\%$) confidence
level. As $\Omega _{m0}$ was sampled with bin size $0.05 $, then the peaks
can be somewhere between $0.00$ and $0.05$.

The joint PDF for the parameter space of $(\Omega _{m0},\Omega _{c0})$, i.e,
$p(\Omega _{m0},\Omega _{c0}\mid \mu _{0})$, is shown in figure \ref
{figOmegam0c0}. The PDF peak is now located at $(\Omega _{m0},\Omega
_{c0})=(0.0,0.81)$, with the same the central value of $\Omega _{m0}$ after
Bayesian marginalization. But, in this two-parameter space with
simultaneously values of $(\Omega _{m0},\Omega _{\Lambda })$, the credible
regions would be given by $(\Omega _{m0},\Omega
_{c0})=(0.0_{-0.00}^{+0.74},0.81_{-0.81}^{+1.31})$ at $1\,\sigma $ ($68,27\%$%
) and $(\Omega _{m0},\Omega
_{c0})=(0.0_{-0.00}^{+1.14},0.81_{-0.81}^{+2.05}) $ at $2\,\sigma $ ($%
95,45\% $) confidence levels. Of course, the integral (marginalization) of $%
p(\Omega _{m0},\Omega _{c0}\mid \mu _{0})$ over the $\Omega _{c0}$
space yields $p(\Omega _{m0}\mid \mu _{0})$, shown in figure
\ref{figsOmegas}. The conclusion is that the marginalization
decreases the confidence regions, even if it does not change the
peak value of probability.

\subsection{The generalized Chaplygin gas density parameter $\Omega _{c0}$}

The generalized Chaplygin gas density parameter $\Omega _{c0}$ estimation
(table \ref{tableParEstGCG}) is like the estimation in the case of CGM shown
by table $5$ of ref. \cite{roberto}, and is quite spread, i.e., it has a
large credible region, for example $\Omega _{c0}=1.39_{-1.25}^{+1.21}$ in
the case of $5$-parameter space. Figure \ref{figsOmegas} show the PDF of $%
\Omega _{c0}$ for the $5$-parameter space (showing $\Omega
_{c0}=1.39_{-0.62}^{+0.59}$ at $1\,\sigma $ confidence level) \ and the $4$%
-parameter space with $k=0$ (when $\Omega _{c0}=1-\Omega _{m0}$).

As discussed above for the GCGM, the joint PDF for the parameter space of $%
(\Omega _{m0},\Omega _{c0})$ in figure \ref{figOmegam0c0} shows \ that the
credible regions would be given by $(\Omega _{m0},\Omega
_{c0})=(0.0_{-0.00}^{+0.74},0.81_{-0.81}^{+1.31})$ at $1\,\sigma $ ($68,27\%$%
) and $(\Omega _{m0},\Omega
_{c0})=(0.0_{-0.00}^{+1.14},0.81_{-0.81}^{+2.05}) $ at $2\,\sigma $ ($%
95,45\% $) confidence level. So, the marginalization of $p(\Omega
_{m0},\Omega _{c0}\mid \mu _{0})$ over the $\Omega _{m0}$ space gives $%
p(\Omega _{c0}\mid \mu _{0})$, yielding a different peak value as well as
different (narrower) confidence regions.

\subsection{The present age $t_{0}$ of the Universe}

The estimates for the dynamical age of the Universe today, $t_{0}$, given in
table \ref{tableParEstGCG}, show that the generalized\ Chaplygin gas model
increases the $2\,\sigma $ credible regions (e.g., $t_{0}=15.3_{-3.2}^{+4.2}%
\,Gy$), when compared to the estimations of the CGM in table $5$ of ref.
\cite{roberto} (e.g., $t_{0}=14.2_{-1.5}^{+2.8}\,Gy$). The peak values of $%
t_{0}$ range between $14.2\,Gy$ to $15.3\,Gy$, while the CGM features $t_{0}$
peaked between $13.1\,Gy$ to $14.8\,Gy$. We expect that a large sample of
SNe Ia will narrow the credible regions and allow, by means of independent
estimations of $t_{0}$ from different observations (globular agglomerates,
etc), to verify which cosmological model is preferred : GCGM, CGM and $%
\Lambda$CDM.

Table \ref{tableBestFitGCG} shows the $t_{0}$ for the best-fitting
parameters without any marginalization (i.e., the $n$-parameter model has
its $n$ parameters taken simultaneously yielding a minimum $\chi ^{2}$) :
the $t_{0}$ ranges from $14.0Gy$ to $14.4\,Gy$, well inside the estimated
credible regions for $t_{0}$.

\subsection{The deceleration parameter $q_{0}$ of the Universe}

All the estimations based on the generalized Chaplygin gas model (table \ref
{tableParEstGCG}) suggest an accelerating Universe today, i.e., $q_{0}<0$,
with a high confidence level ranging from $1.66\,\sigma $ ($90.24\,\%$) to $%
3.72\,\sigma $ ($99.98\,\%$), depending on the curvature and cold dark
matter densities of the Universe. The peak values of $q_{0}$ range between $%
-0.89$ to $-0.59$. The data of CGM in table $5$ of ref. \cite{roberto} show
a similar behaviour.

Although the $q_{0}$\ values of Table \ref{tableBestFitGCG}, from $-1.00$ to
$-0.80$, are inside the estimated credible regions for $q_{0}$, the
estimated peak values of $q_{0}$ differs from the $q_{0}$ for the
best-fitting parameters.

\subsection{The inflexion point $a_{i}$ for the scale factor}

We have also estimated the value of the scale factor that shows the
beginning of the recent accelerating phase of the Universe. Table \ref
{tableParEstGCG} shows that the most probable value for $a_{i}$ ranges
between $0.57\,a_{0}$ to $0.77\,a_{0}$, where $a_{0}$ is the value of scale
factor today, with usually an small upper credible region and a larger lower
credible region.

The probability of $a_{i}<a_{0}$ is theoretically the same having an
accelerating Universe today ($q_{0}<0$), and these independent probability
calculations agree almost exactly as table \ref{tableParEstGCG}\ shows,
indicating that the Bayesian probability analyses of this work are accurate
and reliable for the given SNe Ia data.

We can obviously express the scale factor value $a_{i}$ in terms of redshift
value $z_{i}$, $a_{i}/a_{0}=1/(1+z_{i})$ or $z_{i}=-1+(a_{0}/a_{i})$. For
example, $a_{i}=0.67_{-0.37}^{+0.25}\,a_{0}$ corresponds to $%
z_{i}=0.49_{-0.41}^{+1.87}$. We expect that a large sample of SNe Ia with
large redshift values will help improve this type of estimation of $z_{i}$
or equivalently $a_{i}$.

The $a_{i}$ values are almost equal in table \ref{tableBestFitGCG} for the
best-fitting parameters, between $0.756\,a_{0}$ and $0.792\,a_{0}$, and this
range fits well inside the the estimated credible regions for $a_{i}$ of
table \ref{tableParEstGCG}.

\subsection{The eternally expanding Universe}

The probability of having a eternally expanding Universe, given here by
demanding $\dot{a}>0$ during the future evolution of the Universe, is
estimated to be exactly one or almost one (at almost a $6\,\sigma $ level),
see table \ref{tableParEstGCG}. The estimations for the CGM and $\Lambda$CDM
cases (tables $4$ and $5$, ref. \cite{roberto}) were miscalculated, they are
also exactly one or almost one.

\section{Conclusion}

In this paper we investigated some constraints on the generalized Chaplygin
gas model (GCGM) using type Ia supernovae data and Bayesian statistics. The
model is built with the Chaplygin gas and a pressureless fluid as the matter
content of the Universe, whose density parameters today are $\Omega _{c0}$
and $\Omega _{m0}$ respectively, such that the curvature term is $\Omega
_{k0}=1-\Omega _{m0}-\Omega _{c0}$. The parameter $\bar{A}$ related to the
sound velocity of the Chaplygin gas, the Hubble parameter $H_{0}$ and the $%
\alpha $ parameter are also independent free variables. We allowed the
parameter $\alpha $ to vary between negative and large positive values. In
principle, this could lead to instability problems at perturbative level ($%
\alpha $ negative), or superluminal sound velocity ($\alpha $ greater than
one), but both potential problems can be resolved using a fundamental
description for the GCG.

When the five independent parameters are free, the Bayesian estimation
yields $\alpha =-0.86_{-0.15}^{+6.01}$, $H_{0}=62.0_{-1.42}^{+1.32}%
\,km/Mpc.s $, $\Omega _{k0}=-1.26_{-1.42}^{+1.32}$, $\Omega
_{m0}=0.00_{-0.00}^{+0.86}$, $\Omega _{c0}=1.39_{-1.25}^{+1.21}$, $\bar{A}%
=1.00_{-0.39}^{+0.00}$, $t_{0}=15.3_{-3.2}^{+4.2}$, $%
q_{0}=-0.80_{-0.62}^{+0.86}$ and $a_{i}=0.67_{-0.37}^{+0.25}\,a_{0}$ where $%
t_{0}$ is the age of the Universe, $q_{0}$ is the value of the deceleration
parameter today and $a_{i}$ is the value of the scale factor that marks the
start of the recent accelerating phase of the Universe. We have also
estimated the cosmological parameters fixing the pressureless matter
(through the constraints of nucleosynthesis) or the curvature (imposing the
flat condition), or even both. The estimations of the remaining free
parameters do not change dramatically in each case, with some exceptions.
For example, the prediction for the curvature parameter depends on whenever
the pressureless matter is fixed or not : a spatially flat Universe is
favoured only if $\Omega _{m0}$ is fixed at low values.

In particular, we verify that the $\alpha $ parameter is strongly related to
the $\bar{A}$ parameter. In this manner, the best value of $\alpha $ have
the peak values and the credible regions changed in function of the
marginalization process used. The $\alpha $ estimations are very spread and
differ a lot if one or two-dimensional parameter space is used. Many authors
\cite{martin}--\cite{bertolami2004b}, prefer to estimate $\alpha $ and $\bar{%
A}$ simultaneously using two-dimensional credible regions and they do not
marginalize the PDF to obtain the more statistically robust one-dimensional
PDFs for $\alpha $ and $\bar{A}$, so any comparison should be taken
carefully as the credible region estimations on two-dimensional parameter
space are usually different from the one-dimensional estimations.

Here we take the marginalization approach to obtain
one-dimensional PDFs, and one important point arises in this
analysis : despite the dependency on the $\bar{A}$ parameter, the
value $\alpha =1$, that represents the ordinary CGM, is not ruled
out since this particular case is preferred at $39.3\%$ to
$63.2\%$, depending on the spatial section and matter content.

We calculated the value of the scale factor when the dark energy starts to
prevail, $a_{i}=0.67~a_{0}$, for the case of five free parameters. In
general, the value for $a_{i}$ ranges between $0.57~a_{0}$ and $0.77~a_{0}$,
where $a_{0}$ is the value of the scale factor today. This results are
coherent with the suposition that the recent acceleration of the Universe
could be started with the redshift parameter $z$ between $0$ and $1$
obtained by the type Ia supernovae data.

In the case of the ordinary matter parameter $\Omega _{m0}$ our
analyses show that the preferred values are small, around zero.
This situation is the same of the case of the ordinary Chaplygin
gas. The result reinforces the idea of the Quartessence model
\cite{pasquier, bilic2002, martin}, that unifies dark matter and
dark energy into a single fluid, the Chaplygin gas. The predicted
age of the Universe is well above the age of globular clusters. At
same time, the deceleration parameter is consistent with the
estimations from the WMAP data.

Further developments of the present work include the use of a
large sample of type Ia supernovae. However, the sample of $26$
SNe Ia used here has an excellent quality with respect to the low
values of $\chi _{\nu }^{2}$. Moreover, the dispersion of the
parameter estimations obtained here has the same order of
estimations using larger samples. In some cases our estimation are
even less spread, as our preliminary investigation on larger SNe
Ia indicates.

We also plan to apply the Bayesian analysis to the GCGM using
large scale structures data (2dFRGS) and to the anisotropy of the
cosmic microwave background. In this way, we may consistently
cross the estimations of different observational data and
constrain more sharply the predictions to each free parameter of
the model.

\vspace{0.5cm} \noindent \textbf{Acknowledgments} \newline
\newline
\noindent We would like to thank Orfeu Bertolami, Ioav Waga and
Mart\'{\i}n Makler for many discussions about the generalized
Cha\-ply\-gin gas model. The authors are also grateful by the
received financial support during this work, from CNPq (J. C. F.
and S. V. B. G.) and FACITEC/PMV (R. C. Jr.).

\end{document}